\documentclass{SCIS2024}

\begin{document}

\ArticleType{REVIEW}
\Year{2024}
\Month{}
\Vol{}
\No{}
\DOI{}
\ArtNo{}
\ReceiveDate{}
\ReviseDate{}
\AcceptDate{}
\OnlineDate{}

\title{An Overview on IRS-Enabled Sensing and Communications for 6G: Architectures, Fundamental Limits, and Joint Beamforming Designs}{An Overview on IRS-enabled Sensing and Communications for 6G: Architectures, Fundamental Limits, and Joint Beamforming Designs}

\author[1,2]{Xianxin~SONG}{}
\author[2]{Yuan~FANG}{}
\author[3]{Feng~WANG}{}
\author[1,4]{Zixiang~REN}{}
\author[2]{\\Xianghao~YU}{}
\author[5]{Ye~ZHANG}{{yezhang@bistu.edu.cn}}
\author[6]{Fan~LIU}{}
\author[1]{Jie~XU}{{xujie@cuhk.edu.cn}}
\author[7]{Derrick Wing Kwan NG}{}
\author[8,9]{\\Rui ZHANG}{}
\author[1]{Shuguang CUI}{}
\AuthorMark{Song X X, Fang Y, Wang F}

\AuthorCitation{Song X X, Fang Y, Wang F, et al}



\address[1]{School of Science and Engineering (SSE), Shenzhen Future Network of Intelligence Institute (FNii-Shenzhen),\\ and Guangdong Provincial Key Laboratory of Future Networks of Intelligence,\\ The Chinese University of Hong Kong (Shenzhen), Guangdong {\rm 518172}, China}
\address[2]{Department of Electrical Engineering, City University of
Hong Kong, Hong Kong, China}
\address[3]{School of Information Engineering, Guangdong University of Technology, Guangzhou {\rm 510006}, China}
\address[4]{Key Laboratory of Wireless-Optical Communications, Chinese Academy of Sciences,\\ School of Information Science and Technology, University of Science and Technology of China, Hefei {\rm230027},China}
\address[5]{Computer School, Beijing Information Science and Technology University, Beijing {\rm 100096}, China}
\address[6]{National Mobile Communications Research Laboratory, Southeast University, Nanjing {\rm 210096}, China}
\address[7]{School of Electrical Engineering and Telecommunications, University of New South Wales,\\ Sydney, NSW 2052, Australia}
\address[8]{School of Science and Engineering, Shenzhen Research Institute of Big Data,\\ The Chinese University of Hong Kong, Shenzhen, Guangdong {\rm 518172}, China}
\address[9]{Department of Electrical and Computer Engineering, National University of Singapore, {\rm 117583}, Singapore}
\abstract{This paper presents an overview on intelligent reflecting surface (IRS)-enabled sensing and communication for the forthcoming sixth-generation (6G) wireless networks, in which IRSs are strategically deployed to proactively reconfigure wireless environments to improve both sensing and communication (S\&C) performance. First, we exploit a single IRS to enable wireless sensing in the base station's (BS's) non-line-of-sight (NLoS) area. In particular, we present three IRS-enabled NLoS target sensing architectures with fully-passive, semi-passive, and active IRSs, respectively. We compare their pros and cons by analyzing the fundamental sensing performance limits for target detection and parameter estimation. Next, we consider a single IRS to facilitate integrated sensing and communication (ISAC), in which the transmit signals at the BS are used for achieving both S\&C functionalities, aided by the IRS through reflective beamforming. We present joint transmit signal and receiver processing designs for realizing efficient ISAC, and jointly optimize the transmit beamforming at the BS and reflective beamforming at the IRS to balance the fundamental performance tradeoff between S\&C. Furthermore, we discuss multi-IRS networked ISAC, by particularly focusing on multi-IRS-enabled multi-link ISAC, multi-region ISAC, and ISAC signal routing, respectively. Finally, we highlight various promising research topics in this area to motivate future work.}


\maketitle

\section{Introduction} 
Sixth-generation (6G) wireless networks are evolving from merely connected everything to enhancing connected intelligence, supporting numerous emerging applications such as low-altitude economy, autonomous driving, and industrial automation\cite{huawei20226g,cheng2024networked,10634888}. In particular, integrated sensing and communication (ISAC) has been widely recognized as an essential new usage scenario for 6G wireless networks, in which cellular base stations (BSs) and mobile devices exploit wireless signals for both sensing and communication (S\&C) functionalities \cite{8999605,9737357,9933849,9705498,9606831,9540344,IMT2030}. With the integrated hardware and system designs for S\&C, ISAC can improve spectrum utilization efficiency, reduce energy consumption, and promote S\&C performance through collaborative technologies. Leveraging advanced technologies such as extremely large antenna arrays\cite{9314267,9500972}, millimeter (mmWave)\cite{8373698,8700132}, and terahertz (THz)\cite{9681870,10045774}, 6G networks are expected to provide significantly enhanced sensing resolution and accuracy as well as increased communication data rates and reliabilities. Various prior works \cite{9124713,9246715,9415119,10086626,10153696,9916163,10217169,9652071,10251151,10584287,9791349,hua2024near,10388218} have investigated adaptive transmitter and receiver designs to address the challenges caused by dynamically changing wireless environments. However, due to randomly distributed obstacles and scatterers in environments, the wireless channels may suffer from shadowing and deep fading, and the line-of-sight (LoS) links between the BS and the communication users (CUs)/targets may be blocked, seriously degrading both S\&C performances. Although blocked targets may be sensed by through-wall radar (TWR) \cite{10.1145/2816795.2818072} and looking around corner (LAC)\cite{7272840} technologies, their effectiveness highly relies on appropriate materials of obstacles and static transmission environment. As such, traditional ISAC transceiver designs \cite{9124713,9246715,9415119,10086626,10153696,9916163,10217169,9652071,10251151,10584287,9791349,hua2024near,10388218} may not be able to provide ubiquitous S\&C services over dynamic and complex channel environments.

Intelligent reflecting surface (IRS) \cite{8910627,9326394,9690635,8811733,9122596,9483903,10555049,10680462} or reconfigurable intelligence surface (RIS) \cite{9140329,9424177,9847080}  has been recognized as a cost-effective technology to address the aforementioned issues faced in conventional ISAC systems. An IRS is a metasurface consisting of numerous low-cost reflecting elements capable of inducing certain phase shifts and/or amplifications to incident signals. With controllable signal reflection, IRS can proactively reconfigure the wireless transmission channel between transceivers, for enhancing S\&C performances. Specifically, for wireless communications, IRS can potentially increase the channel rank between transceivers, enhance signal power, and mitigate the undesired inter-user interference through controllable reflective beamforming design\cite{8910627,9326394,9690635,8811733,9122596,9483903,10555049,10680462,9365004,10159991,9847080,9140329,9424177,9241752,10159017,10643789,9427474,9586067,9963672}. For wireless sensing, the controllable IRS reflection not only helps sensing signals bypass environmental obstacles to enable non-line-of-sight (NLoS) sensing, but also supports advanced multi-angle sensing to provide more comprehensive target information in addition to conventional LoS sensing\cite{10008725,10138058,10464564,xianxin_comparsion,9508883,9454375,9732186,10422881,9540372,10103813,9625826,10113892,9724202,active_IRS_xianxin,fang2024joint,10443321}. For ISAC, IRS can also support adaptive interference management to efficiently control the mutual interference between S\&C\cite{9771801,10077119,10243495,10702570,10279464,10440056,10497119,10197455,9416177,9364358,10254508,9769997,9729741,9591331,10086570,10050406,9979782,10054402,10319318,10496515,10184278,10186271,10149664,10226306,10304580}.

\begin{figure}[h]
        \centering
        \includegraphics[width=0.75\textwidth]{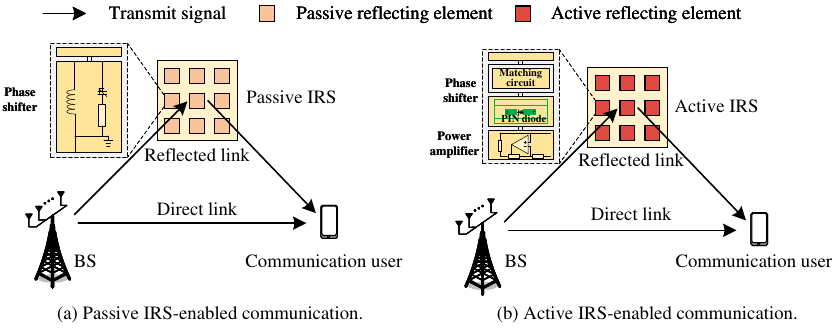}
        \caption{Basic architectures of IRS-enabled wireless communication.}
        \label{IRS_enabled_communication_architectures}
\end{figure}
\begin{figure}[h]
        \centering
        \includegraphics[width=1.0\textwidth]{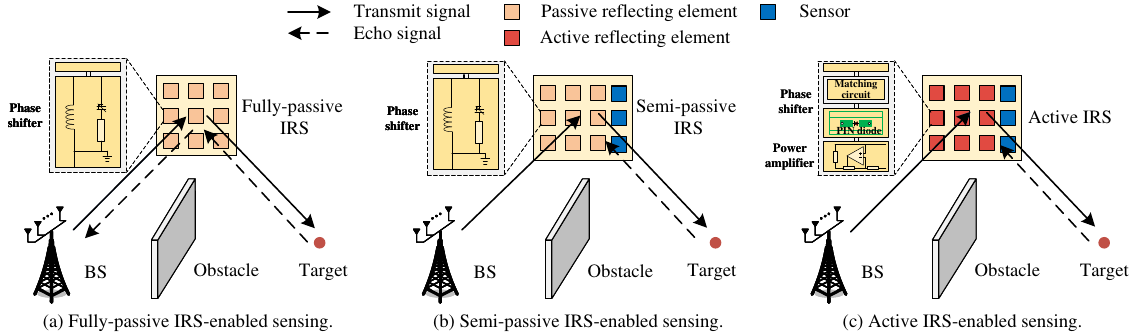}
        \caption{Basic architectures of IRS-enabled wireless sensing.}
        \label{IRS_enabled_sensing_architectures}
\end{figure}

Despite the aforementioned advancements, IRS-enabled S\&C still face significant challenges. 
\begin{itemize}
\item First, evaluating the fundamental performance limits of IRS-enabled S\&C systems is essential but challenging. For wireless communication systems, according to whether the IRS has the ability to alter the amplitude of incident signals, passive and active IRS-enabled communication are two typical architectures, as shown in Figure~\ref{IRS_enabled_communication_architectures}. Prior works \cite{8910627,9326394,9690635,8811733,9122596,9483903,10555049,10680462,9365004,10159991,9847080,9140329,9424177,9241752,10159017,10643789,9427474,9586067,9963672} have explored the communication performance limits using various performance metrics, such as communication signal-to-noise ratio (SNR) and channel capacity. For wireless sensing system, as shown in Figure~\ref{IRS_enabled_sensing_architectures}, there are three typical IRS-enabled NLoS wireless sensing architectures in general, namely fully-passive, semi-passive, and active IRS-enabled sensing architectures, according to the types of reflecting elements and the deployment of sensors\cite{10008725,10138058,10464564,xianxin_comparsion,9508883,9454375,9732186,10422881,9540372,10103813,9625826,10113892,9724202,active_IRS_xianxin,fang2024joint}. For various sensing tasks such as detection and estimation, the characterization of fundamental sensing performance limits across different architectures is still unclear.

\item Second, as ISAC integrates both S\&C functionalities into a system, it necessitates a redesign of the transmitted signals to support both services and balance the performance tradeoff between them. On the one hand, the transmit signal design criteria for S\&C functions are fundamentally different. For communication, the BSs aim to transmit messages, unknown to the CUs, as reliably as possible\cite{6773024}. By contrast, for wireless sensing, the BSs intend to acquire precise knowledge about the environment, such as the presence or the motion parameters of targets.\cite{kay1993fundamentals,kay1993fundamentalsdetection,richards2014fundamentals}. On the other hand, it is necessary to explore the ISAC performance tradeoff limits by jointly optimizing the transmit and reflective beamforming design, which is challenging due to the non-convex nature of the resultant optimization problems and the coupled relation between the transmit and reflective beamforming design. 

\item Third, considering the limited sensing region and resolution of a single IRS-enabled sensing system, multi-IRS networked S\&C are emerging\cite{9241752,10159017,10643789,10497119,fang2024joint}, which can significantly enhance S\&C performance by seamlessly fusing and sharing sensory data among multiple IRSs. In this scenario, however, the signals reflected by multiple IRSs are superimposed with each other. Thus, managing the interference and extracting the interested sensing results from the superimposed received signals are challenging. Meanwhile, the sensing data fusion and sharing among multiple IRSs may result in considerably huge energy consumption and communication overheads.
\end{itemize}

This paper aims to provide an overview of the recent advancements in IRS-enabled S\&C in addressing the above problems. First, we present three IRS-enabled sensing architectures with fully-passive, semi-passive, and active IRSs, respectively, according to whether the IRS is mounted with passive or active reflecting elements and whether it is deployed with sensors. We consider the target detection and parameter estimation tasks, and accordingly explore the sensing performance limits in detection probability and estimation Cram\'er-Rao bound (CRB). Next, we adopt the IRS to simultaneously facilitate S\&C in an ISAC system. In particular, we develop a new joint signal design tailored to the distinct criteria of S\&C and introduce a signal interference cancellation method at the CU. Then, we jointly optimize the transmit and reflective beamforming design to explore the ISAC performance boundaries. Furthermore, we introduce the multi-IRS networked S\&C, which can significantly extend the ISAC region and enhance the ISAC performance through  collaboration among multiple IRSs. Finally, we discuss some interesting open problems along this direction, such as  sensing-assisted IRS-enabled communication, IRS deployment problem for sensing and ISAC, mounting IRS at target for facilitating sensing performance and privacy, near-field IRS sensing and ISAC, IRS-enabled wideband ISAC, and machine learning for IRS-enabled ISAC design.

In the literature, there have been some prior works on the overview about IRS-enabled sensing\cite{10422881} and ISAC\cite{10243495,10077119,10702570}. For instance, \cite{10422881} reviewed the IRS-enabled sensing architectures and design issues with fully-passive, semi-passive, and active IRSs. However, this work lacked a comprehensive comparison of the sensing performance limits in detection and estimation tasks with various IRSs. Furthermore, \cite{10077119} introduced the great potentials of fully-passive IRS for sensing and ISAC, which, however, only focused on the case with a single fully-passive IRS and lacked the comparison between multiple types of IRSs. In addition, \cite{10243495} analyzed the architectures and benefits of fully-passive IRS for sensing and ISAC from the signal processing perspective. However, this work only focused on the case with a single IRS and lacked the analysis of semi-passive IRS and active IRS for sensing and ISAC. Besides, in \cite{10702570}, the authors reviewed IRS-enabled sensing by focusing on sensing channel characteristics, IRS architectures, resources allocation and cooperation between S\&C. However, this work \cite{10702570} did not discuss the fundamental sensing performance limits of IRS-enabled sensing in detection and estimation tasks, and also lacked the analysis about the transmit signal and joint beamforming designs. Different from prior works, this paper provides a more comprehensive overview of IRS-enabled S\&C by focusing on the detection and estimation performance characterization with fully-passive, semi-passive, and active IRSs, joint signal design and performance tradeoff for single-IRS-enabled ISAC, and multi-IRS networked S\&C.

\textit{Notations:} 
Boldface  lower-case and boldface capital letters refer to vectors and matrices, respectively. $\mathbb{R}^{x \times y}$ and $\mathbb{C}^{x \times y}$ denote the spaces of real and complex  matrices with dimension $x \times y$, respectively. $\mathrm {tr}(\mathbf A)$ and $\mathrm{rank}(\mathbf A)$ denote the trace and rank of matrix $\mathbf A$, respectively. $\mathrm{Re}(\cdot)$ and $\mathrm{Im}(\cdot)$ denote the real and imaginary parts of the argument, respectively. $\mathbf A^*$, $\mathbf A^{T}$,  $\mathbf A^{H}$, and $\mathbf A^{-1}$ denote the conjugate, transpose, conjugate transpose, and inverse of matrix $\mathbf A$, respectively. $\mathrm {diag}(a_1,\cdots,a_N)$ denotes a diagonal matrix with diagonal elements $a_1,\cdots,a_N$. $\mathbf A \succeq \mathbf{0}$ means that matrix $\mathbf A$ is positive semi-definite. $\mathbb{E}(\cdot)$ denotes the expectation operation. $\|\cdot\|$ denotes the Euclidean norm. $|\cdot|$ denotes the absolute value of a complex scalar. $\otimes$ denotes the Kronecker product. For vector $\mathbf x$, $ \mathrm {arg}(\mathbf x)$ corresponds to a vector with each element being the phase of the corresponding element in $\mathbf x$. The imaginary unit is denoted as $j=\sqrt{-1}$. Function $Q_m(a,b)=\frac{1}{a^{m-1}}\int_{b}^\infty x^m e^{-\frac{x^2+a^2}{2}}I_{m-1}(ax)dx$ is the generalized Marcum Q-function of order $m$ for non-centrality parameter $a$, in which $I_m (\cdot)$ denotes the modified Bessel function of the first kind of order $m$. Furthermore, we define  $\mathcal{Q}_{\chi^2_{\nu}(\lambda)}(x) = \int_{x}^{\infty}p(x)dx,x>0$,  with $p(x) = \begin{cases}
\frac{1}{2}\left(\frac{x}{\lambda}\right)^{\frac{\nu-2}{4}}\exp\left[-\frac{1}{2}(x+\lambda)\right] I_{\frac{\nu}{2}-1}(\sqrt{\lambda x}), & x>0,\\
0, &x<0.
\end{cases}$

\section{IRS-Enabled Wireless Sensing}\label{sec:sensing}

This section considers an IRS-enabled sensing system, by focusing on the basic architectures and fundamental sensing performance limits. In this system,  an IRS is deployed to extend the sensing coverage by reflecting the transmit signal from the BS to bypass obstacles. The deployment of IRS provides the possibility for NLoS target sensing, but introduces new technical challenges in sensing architecture design and performance evaluation. Among various target sensing tasks, we focus on target detection and parameter estimation, which aim to determine the presence or absence of the target and estimate its parameters, respectively.

\subsection{Architectures for IRS-Enabled Sensing}
As shown in Figure~\ref{IRS_enabled_sensing_architectures}, there are three typical architectures of IRS-enabled NLoS wireless sensing, namely fully-passive, semi-passive, and active IRS-enabled sensing architectures, respectively\cite{10008725,10138058,10464564,xianxin_comparsion,9508883,9454375,9732186,10422881,9724202,9540372,10103813,9625826,10113892,active_IRS_xianxin}. The three architectures differ in whether the IRS is equipped with passive reflecting elements with only phase alteration ability or active reflecting elements with both phase alteration and signal power amplification abilities, and whether the sensing is performed at the IRS exploiting dedicated sensors or at the BS. In the following, we first introduce the system model and then analyze their advantages and drawbacks.

First, we consider the fully-passive IRS-enabled sensing system in Figure~\ref{IRS_enabled_sensing_architectures}(a). The fully-passive IRS only consists of several passive reflecting elements that can alter the phase of incident signals\cite{10008725,10138058,10464564,xianxin_comparsion,9508883,9454375,9732186,10422881,9540372,10103813,9625826,10113892}. Let $M_t$ denote the number of transmit antennas at the BS. Let $M_r$ denote the number of receive antennas for sensing at the BS. Let $\mathcal T\triangleq\{1,\cdots,T\}$ denote the set of sensing symbols with $T$ being the number of symbols. Let $\mathbf x(t) \in \mathbb{C}^{M_t\times 1}$ denote the transmit signal at sensing symbol $t$ with sample covariance matrix $\mathbf R =\frac{1}{T}\sum_{t=1}^T\mathbf x(t)\mathbf x^H(t)$. Let $P_\text{BS}$ represent the maximum transmit power at the BS. The sample covariance matrix should satisfy $\mathbf R \succeq \mathbf{0}$ and $\mathrm{tr}(\mathbf R) \le P_\text{BS}$. Let $\mathcal{N}\triangleq\{1,\cdots,N\}$ denote the set of reflecting elements with $N$ being the number of reflecting elements at the IRS. Let matrices $\mathbf G_t\in \mathbb{C}^{N\times M_t}$ and $\mathbf G_r\in \mathbb{C}^{M_r\times N}$ denote the channel from the BS to the IRS and that from the IRS to the BS, respectively. Let $\mathbf \Phi= \mathrm{diag}(a_1e^{j\phi_1},\ldots,a_N e^{j\phi_{N}})\in \mathbb{C}^{N\times N}$ represent the IRS reflecting matrix, where $\phi_{n}\in(0,2\pi]$ and $0 \le a_n\le a_{\max}, n \in \mathcal N$, denote the phase shift and amplification gain introduced by reflecting element $n$, respectively, with $a_{\max}$ being the maximum amplification gain at each reflecting element. For the fully-passive IRS, the amplification gain at each reflecting element is unity, i.e., $a_n=1, n\in\mathcal N$. Thus, the reflecting matrix is further modeled as $\mathbf \Phi= \mathrm{diag}(e^{j\phi_1},\ldots,e^{j\phi_{N}})\in \mathbb{C}^{N\times N}$. Let $\mathbf H_1 \in \mathbb C^{N\times N}$ denote the target response matrix with respect to (w.r.t.) the fully-passive IRS. When the sensing target is a point target, the target response matrix is modeled as $\mathbf H_1 = \alpha\mathbf a (\theta)\mathbf a^{T}(\theta)$, where $\alpha\in\mathbb C$ is the channel coefficient of the IRS-target-IRS link, $\theta$ is the angle of the target w.r.t. the IRS, and $\mathbf a(\theta)\in\mathbb C^{N\times 1}$ is the steering vector of IRS reflecting elements toward angle $\theta$, respectively. 
As the fully-passive IRS lacks the ability to receive the target's echo signals, the BS performs target sensing by processing the echo signals traveling through the BS-IRS-target-IRS-BS link. The received signal by the BS is
\begin{equation}\label{eq:echo_fully_passive}
\mathbf y_1(t) = \mathbf G_r\mathbf \Phi^T\mathbf H_1\mathbf\Phi\mathbf G_t\mathbf x(t) + \mathbf n_1(t),~t \in \mathcal T,
\end{equation}
where $\mathbf n_1(t)\sim \mathcal{C N}(\mathbf{0}, \sigma^2\mathbf I_{M_r})$ denotes the noise at the BS's receiver with $\sigma^2$ being the noise power at each receive antenna. Based on the received signal in \eqref{eq:echo_fully_passive}, the sensing SNR with fully-passive IRS is 
\begin{equation}
\text{SNR}_1 = \frac{\mathbb E(\|\mathbf G_r\mathbf \Phi^T\mathbf H_1\mathbf\Phi\mathbf G_t\mathbf x(t)\|^2)}{\mathbb E(\|\mathbf n_1(t)\|^2)}=
\frac{\mathrm{tr}\left(\mathbf G_r\mathbf \Phi^T\mathbf H_1\mathbf\Phi\mathbf G_t\mathbf R\mathbf G_t^H\mathbf \Phi^H\mathbf H_1^H\mathbf \Phi^*\mathbf G_r^H\right)}{\sigma^2}.
\end{equation}

Next, we introduce the semi-passive IRS-enabled sensing system in Figure~\ref{IRS_enabled_sensing_architectures}(b). Compared with fully-passive IRS, semi-passive IRS is further mounted with dedicated sensors for receiving the target's echo signals\cite{10464564,xianxin_comparsion,10422881,9724202}. We assume that the number of sensors at the semi-passive IRS is the same as that of receive antennas at the BS for fully-passive IRS-enabled sensing, for a fair comparison. Note that for semi-passive IRS, we have $a_n=1, n \in\mathcal N$, and thus the IRS reflecting matrix is modeled as $\mathbf \Phi= \mathrm{diag}(e^{j\phi_1},\ldots,e^{j\phi_{N}})\in \mathbb{C}^{N\times N}$. Let $\mathbf H_2 \in \mathbb C^{M_r\times N}$ denote the target response matrix w.r.t. the semi-passive IRS, which is further modeled as $\mathbf H_2 = \alpha\mathbf b(\theta)\mathbf a^{T}(\theta)$ for the point target, where $\mathbf b(\theta)\in\mathbb C^{M_r\times 1}$ is the steering vector of sensors toward target's angle $\theta$. In this case, the received signal by the IRS's sensors through the BS-IRS-target-IRS link is  
\begin{equation}\label{echo_semi_IRS}
\mathbf y_2(t) = \mathbf H_2\mathbf\Phi\mathbf G_t\mathbf x(t) + \mathbf n_2(t), ~t \in \mathcal T,
\end{equation}
where $\mathbf n_2(t) \sim \mathcal{C N}(\mathbf{0}, \sigma^2\mathbf I_{M_r})$ is the noise at the IRS's sensors with $\sigma^2$ being the noise power at each sensor. The sensing SNR at the IRS's sensor is 
\begin{equation}\label{eq:SNR_semi}
\text{SNR}_2 =\frac{\mathbb E(\|\mathbf H_2\mathbf\Phi\mathbf G_t\mathbf x(t)\|^2)}{\mathbb E(\|\mathbf n_2(t)\|^2)}=\frac{\mathrm{tr}\left(\mathbf H_2\mathbf\Phi\mathbf G_t\mathbf R\mathbf G_t^H\mathbf \Phi^H\mathbf H_2^H\right)}{\sigma^2}.
\end{equation}

Furthermore, we consider the active IRS-enabled sensing system in Figure~\ref{IRS_enabled_sensing_architectures}(c)\cite{active_IRS_xianxin,fang2024joint}. The active IRS is composed of multiple active reflecting elements, each of which can alter the phase and amplify the power of incident signals. In active IRS, the amplification gain at each reflecting elements can be higher than one. The active IRS also employs dedicated sensors for receiving the target's echo signals, similarly as for semi-passive IRS. Different from fully- and semi-passive IRSs, the signal reflection at the active IRS induces additional noise to the incident signal due to the amplifier. The target response matrix w.r.t the active IRS is same as that for the semi-passive IRS, i.e., $\mathbf H_2$. The received echo signal by the IRS's sensors through the BS-IRS-target-IRS link is given as 
\begin{equation}\label{echo_active_IRS}
\mathbf y_3(t) = \mathbf H_2\mathbf\Phi\mathbf G_t\mathbf x(t) +\mathbf H_2\mathbf\Phi\mathbf z(t) + \mathbf n_3(t), ~t \in \mathcal T,
\end{equation}
where $\mathbf z(t)\sim \mathcal{C N}(\mathbf{0}, \sigma_z^2\mathbf I_{N})$ is the reflection noise due to signal power amplification with $\sigma_z^2$  being the power of noise at each reflecting element and $\mathbf n_3(t)\sim \mathcal{C N}(\mathbf{0}, \sigma^2\mathbf I_{M_r})$ is the noise at the IRS's sensors. Let $P_\text{IRS}$ denote the maximum power at the active IRS. The power constraint at the active IRS is modeled as $\mathbb E(\|\mathbf\Phi\mathbf G_t\mathbf x(t)\|^2)  + \mathbb E(\|\mathbf\Phi\mathbf z(t)\|^2)=\mathrm{tr}(\mathbf \Phi\mathbf G_t\mathbf R\mathbf G_t^H \mathbf \Phi^H) + \sigma_z^2\mathrm{tr}(\mathbf \Phi\mathbf \Phi^H)\le P_\text{IRS}$. The sensing SNR with active IRS is  
\begin{equation}\label{eq:SNR_active}
\text{SNR}_3=\frac{\mathbb E(\|\mathbf H_2\mathbf\Phi\mathbf G_t\mathbf x(t)\|^2)}{\mathbb E(\|\mathbf H_2\mathbf\Phi\mathbf z(t)\|^2+\|\mathbf n_3(t)\|^2)}=\frac{\mathrm{tr}\left(\mathbf H_2\mathbf\Phi\mathbf G_t\mathbf R\mathbf G_t^H\mathbf \Phi^H\mathbf H_2^H\right)}{\mathrm{tr}\left(\mathbf \Phi^H\mathbf H_2^H\mathbf H_2\mathbf\Phi\right)\sigma_z^2+\sigma^2}.
\end{equation}

The advantages and drawbacks of different IRS-enabled sensing architectures are given by comparing the signal models in \eqref{eq:echo_fully_passive}, \eqref{echo_semi_IRS}, and \eqref{echo_active_IRS}, as well as the hardware components with different types of IRSs. 
\begin{itemize}
\item For fully-passive IRS sensing, the target sensing is performed at the BS, and thus the received signal with fully-passive IRS suffers from an additional path loss over the IRS-BS channel  $\mathbf G_r$, compared to semi-passive and active IRSs deployed with sensors for receiving sensing signal. In general, additional path loss is detrimental to sensing performance. However, this extra signal reflection of the IRS-BS link also benefits from an additional reflective beamforming gain. 
It is shown in \cite{xianxin_comparsion} that for the far-filed target sensing when the number of reflecting elements is sufficiently large and with proper reflective beamforming design, the additional reflective beamforming gain can surpass the additional path loss, thus making the sensing SNR with a fully-passive IRS outperform that with a semi-passive IRS. Another benefit of  fully-passive IRS is its low hardware cost and energy consumption without deploying sensors and amplifiers at the IRS, as compared to the other two architectures. 

\item Semi-passive IRS-enabled sensing experiences a smaller path loss than its fully-passive counterpart, but experiences less reflective beamforming gain in the return link. Thus, the sensing performance with a semi-passive IRS can outperform the fully-passive when the distance between the BS and the IRS is large\cite{xianxin_comparsion}. Meanwhile, as the target's echo signals are received by the sensors deployed at the IRS, the sensing data needs to be processed at the IRS or forwarded to the BS for remote signal processing. The hardware cost and system complexity of semi-passive IRS are less than active IRS but higher than fully-passive IRS. 

\item For active IRS-enabled sensing, the reflection-type amplifiers equipped at the IRS can amplify the incident signals to compensate for the distance-product path loss and enhance the signal power towards the sensing target. However, active IRS suffers from higher power consumption and hardware cost. Furthermore, active-IRS introduces additional noise in the reflection procedure, which is in sharp contrast to the other two architectures with noise-free reflecting elements. This may decrease the S\&C performance. Thus, more advanced sensing signal processing algorithms are needed to achieve enhanced sensing performance by mitigating the extra reflection noise at the active IRS\cite{active_IRS_xianxin}.
\end{itemize}

\subsection{Target Detection}
Then, we consider the specific target detection task in the scenario with a point target located at the NLoS region of the BS. In this task, target detection probability and false alarm probability are two fundamental performance metrics\cite{kay1993fundamentalsdetection,richards2014fundamentals}. The target detection probability refers to the likelihood of correctly declaring a target when it is present, while the false alarm probability indicates the likelihood of declaring a target when it is actually absent. 

\subsubsection{Fully-Passive IRS-Enabled Target Detection}
First, we consider the target detection in fully-passive IRS-enabled sensing system with a point target at the NLoS of the BS\cite{xianxin_comparsion}. Let $\mathcal H_0$ and $\mathcal H_1$ denote the hypotheses when the target is absent and present, respectively. By collecting the transmit signals $\mathbf x(t)$, receive signals $\mathbf y_1(t)$, and noise $\mathbf n_1(t)$ in the whole $T$ sensing symbols, we have $\tilde{\mathbf x}=[\mathbf x^T(1),\cdots,\mathbf x^T(T)]^T$, $\tilde{\mathbf y}_1=[\mathbf y_1^T(1),\cdots,\mathbf y_1^T(T)]^T$, and $\tilde{\mathbf n}_1=[\mathbf n_1^T(1),\cdots,\mathbf n_1^T(T)]^T$. Under hypotheses $\mathcal H_0$ and $\mathcal H_1$, the received echo signals at the BS are respectively given as
\begin{subequations}
  \begin{align}
&\mathcal H_0: \tilde{\mathbf y}_1=\tilde{\mathbf n}_1,\\\label{eq:echo_signal_piont_fully}
&\mathcal H_1: \tilde{\mathbf y}_1= \left(\mathbf I_T \otimes\mathbf G_r\mathbf \Phi^T\alpha\mathbf a (\theta)\mathbf a^{T}(\theta)\mathbf\Phi\mathbf G_t\right)\tilde{\mathbf x} + \tilde{\mathbf n}_1.
  \end{align}
\end{subequations}
Then, the probability density functions (PDFs) of the received signal $\tilde{\mathbf y}_1$ under  hypotheses $\mathcal H_0$ and $\mathcal H_1$ are respectively given as
\begin{subequations}
  \begin{align}
&p(\tilde{\mathbf y}_1|\mathcal H_0)=\frac{\exp\left(-\frac{1}{\sigma^2}\|\tilde{\mathbf y}_1\|^2\right)}{\pi^{M_tT}\sigma^2},\\
&p(\tilde{\mathbf y}_1|\mathcal H_1)=\frac{\exp\left(-\frac{1}{\sigma^2}\|\tilde{\mathbf y}_1-\left(\mathbf I_T \otimes\mathbf G_r\mathbf \Phi^T\alpha\mathbf a (\theta)\mathbf a^{T}(\theta)\mathbf\Phi\mathbf G_t\right)\tilde{\mathbf x}\|^2\right)}{\pi^{M_tT}\sigma^2}.
  \end{align}
\end{subequations}
Let $P_{1,\text{D}}$ and $P_{1,\text{FA}}$ denote the detection probability and false alarm probability, respectively. Let $\mathcal C_1\in\mathbb C^{M_rT\times1}$ denote the region  that maps the received signal $\tilde{\mathbf y}_1$ into hypothesis $\mathcal H_1$. Then, we have\cite{kay1993fundamentalsdetection,richards2014fundamentals}
\begin{subequations}
  \begin{align}
P_{1,\text{D}} &= \int_{\tilde{\mathbf y}_1 \in \mathcal C_1}p(\tilde{\mathbf y}_1|\mathcal H_1)d\tilde{\mathbf y}_1,\\
P_{1,\text{FA}} &= \int_{\tilde{\mathbf y}_1 \in \mathcal C_1}p(\tilde{\mathbf y}_1|\mathcal H_0)d\tilde{\mathbf y}_1.
  \end{align}
\end{subequations}

We aim to obtain the detection threshold or the equivalent detection region that maximizes the detection probability under a predetermined false alarm probability\cite{kay1993fundamentalsdetection,richards2014fundamentals}. According to whether the target's parameters are known or unknown, there are  generally two detection models. In the first model, it is assumed that target's parameters, such as the channel coefficient $\alpha$ and angle $\theta$, are known. This model corresponds to the case when the location of target is fixed, and thus we only need to decide the presence of the target. In this case, we employ the Neyman-Pearson (NP) criterion\cite{kay1993fundamentalsdetection,richards2014fundamentals} to determine the optimal detection threshold by comparing the likelihood probabilities under both hypotheses. In the second model, some parameters of the target are unknown. In this case, generalized likelihood ratio test (GLRT) method is used to obtain the detection threshold\cite{kay1993fundamentalsdetection,richards2014fundamentals}. In particular, we first estimate the unknown parameters using maximum likelihood estimation (MLE)\cite{kay1993fundamentals}. Subsequently, we replace the unknown parameters in likelihood functions by the estimated values. Finally, the detection threshold is determined employing the NP criterion, by comparing the the likelihood functions under both hypotheses using the reconstructed likelihood functions. 

In \cite{xianxin_comparsion}, the authors studied the fully-passive IRS-enabled NLoS target detection with a point target located at the NLoS region of the BS, in which the target's angle $\theta$ and the channel coefficient $\alpha$ are assumed unknown. Under such consideration, the GLRT method is adopted to decide the presence of target. The MLEs of $\theta$ and $\alpha$ are respectively given as\cite{10138058,xianxin_comparsion}
\begin{subequations}
  \begin{align}
\theta_{1,\text{MLE}}&= \arg\max_\theta \frac{|\tilde{\mathbf y}_1^H\left(\mathbf I_T \otimes\mathbf G_r\mathbf \Phi^T\mathbf a(\theta)\mathbf a^T(\theta)\mathbf\Phi\mathbf G_t\right)\tilde{\mathbf x}|^2}{\|\left(\mathbf I_T \otimes\mathbf G_r\mathbf \Phi^T\mathbf a(\theta)\mathbf a^T(\theta)\mathbf\Phi\mathbf G_t\right)\tilde{\mathbf x}\|^2},\\
\alpha_{1,\text{MLE}}&=\frac{\tilde{\mathbf y}_1^H\left(\mathbf I_T \otimes\mathbf G_r\mathbf \Phi^T\mathbf a(\theta_{1,\text{MLE}})\mathbf a^T(\theta_{1,\text{MLE}})\mathbf\Phi\mathbf G_t\right)\tilde{\mathbf x}}{\|\left(\mathbf I_T \otimes\mathbf G_r\mathbf \Phi^T\mathbf a(\theta_{1,\text{MLE}})\mathbf a^T(\theta_{1,\text{MLE}})\mathbf\Phi\mathbf G_t\right)\tilde{\mathbf x}\|^2}.
  \end{align}
\end{subequations}
By approximating the actual values of $\theta$ and $\alpha$ with the estimated values and applying the NP criterion, we decide hypothesis $\mathcal H_1$ if 
\begin{equation}\label{eq:detector_1}
L(\tilde{\mathbf y}_1) = \frac{p(\tilde{\mathbf y}_1|\mathcal H_1)}{p(\tilde{\mathbf y}_1|\mathcal H_0)}> \delta_1.
\end{equation}
Otherwise, we decide hypothesis $\mathcal H_0$\cite{kay1993fundamentalsdetection,richards2014fundamentals}. In \eqref{eq:detector_1}, the detection threshold $\delta$ is obtained from the predetermined false alarm probability, i.e., $P_{1,\text{FA}} = \int_{\tilde{\mathbf y}_1:L(\tilde{\mathbf y}_1)> \delta_1}p(\tilde{\mathbf y}_1|\mathcal H_0)d\tilde{\mathbf y}_1$. With proper manipulations, \eqref{eq:detector_1} is equivalent to
\begin{equation}
\frac{|\tilde{\mathbf y}_1^H\left(\mathbf I_T \otimes\mathbf G_r\mathbf \Phi^T\mathbf a(\theta_{1,\text{MLE}})\mathbf a^T(\theta_{1,\text{MLE}})\mathbf\Phi\mathbf G_t\right)\tilde{\mathbf x}|^2}{\|\left(\mathbf I_T \otimes\mathbf G_r\mathbf \Phi^T\mathbf a(\theta_{1,\text{MLE}})\mathbf a^T(\theta_{1,\text{MLE}})\mathbf\Phi\mathbf G_t\right)\tilde{\mathbf x}\|^2} \stackrel{\mathcal{H}_1}{\underset{\mathcal{H}_0}{\gtrless}} \sigma^2\ln\delta_1.
\end{equation}
Through analyzing the distributions of $\frac{|\tilde{\mathbf y}_1^H\left(\mathbf I_T \otimes\mathbf G_r\mathbf \Phi^T\mathbf a(\theta_{1,\text{MLE}})\mathbf a^T(\theta_{1,\text{MLE}})\mathbf\Phi\mathbf G_t\right)\tilde{\mathbf x}|^2}{\|\left(\mathbf I_T \otimes\mathbf G_r\mathbf \Phi^T\mathbf a(\theta_{1,\text{MLE}})\mathbf a^T(\theta_{1,\text{MLE}})\mathbf\Phi\mathbf G_t\right)\tilde{\mathbf x}\|^2}$, the false alarm and detection probabilities are respectively given as \cite{xianxin_comparsion}
\begin{subequations}
  \begin{align}
 P_{1,\text{FA}}&= \frac{1}{\delta_1},\\ 
P_{1,\text{D}}& \approx Q_1\left(\sqrt{2T\text{SNR}_1},\sqrt{2\ln\delta_1}\right)\\\label{eq:PD_FA_fully}
&=Q_1\left(\sqrt{2T\text{SNR}_1},\sqrt{2\ln\frac{1}{P_{1,\text{FA}}}}\right),
  \end{align}
\end{subequations}
where the approximation is due to that the actual value of $\theta$ is approximated by its estimated value, which is valid in the high SNR region\cite{9454375,9732186}. Based on \eqref{eq:PD_FA_fully}, in the fully-passive IRS-enabled sensing system, increasing the sensing SNR is equivalent to increasing detection probability.

\subsubsection{Semi-Passive IRS-Enabled Target Detection}
Next, we introduce the target detection in the semi-passive IRS-enabled sensing system with a point target located at the NLoS region of the BS\cite{xianxin_comparsion}. We also suppose that the target's parameters, i.e., $\theta$ and $\alpha$, are unknown. Similar for the case with fully-passive IRS, the detector based on GLRT is 
\begin{equation}\label{eq:detector_2}
\frac{|\tilde{\mathbf y}_2^H\left(\mathbf I_T \otimes \mathbf b(\theta_{2,\text{MLE}})\mathbf a^{T}(\theta_{2,\text{MLE}})\mathbf\Phi\mathbf G_t\right)\tilde{\mathbf x}|^2}{\|\left(\mathbf I_T \otimes \mathbf b(\theta_{2,\text{MLE}})\mathbf a^{T}(\theta_{2,\text{MLE}})\mathbf\Phi\mathbf G_t\right)\tilde{\mathbf x}\|^2} \stackrel{\mathcal{H}_1}{\underset{\mathcal{H}_0}{\gtrless}}  \sigma^2\ln\delta_2,
\end{equation}
where
\begin{subequations}
  \begin{align}
\theta_{2,\text{MLE}}&= \arg\max_\theta \frac{|\tilde{\mathbf y}_2^H\left(\mathbf I_T \otimes \mathbf b(\theta)\mathbf a^{T}(\theta)\mathbf\Phi\mathbf G_t\right)\tilde{\mathbf x}|^2}{\|\left(\mathbf I_T \otimes \mathbf b(\theta)\mathbf a^{T}(\theta)\mathbf\Phi\mathbf G_t\right)\tilde{\mathbf x}\|^2}.
  \end{align}
\end{subequations}
Let $P_{2,\text{D}}$ and $P_{2,\text{FA}}$ denote the detection probability and false alarm probability with semi-passive IRS, respectively. Similar for the derivation with fully-passive IRS, $P_{2,\text{D}}$ and $P_{2,\text{FA}}$ are respectively given as \cite{xianxin_comparsion}
\begin{subequations}
  \begin{align}
 P_{2,\text{FA}}&= \frac{1}{\delta_2},\\
P_{2,\text{D}}& \approx Q_1\left(\sqrt{2T\text{SNR}_2},\sqrt{2\ln\delta_2}\right)\\ \label{eq:PD_PF_active}
&=Q_1\left(\sqrt{2T\text{SNR}_2},\sqrt{2\ln\frac{1}{P_{2,\text{FA}}}}\right).
  \end{align}
\end{subequations}
Note that the detection probability in \eqref{eq:PD_PF_active} is increased with an increase in the sensing SNR, similarly to the fully-passive IRS case. 

\subsubsection{Active IRS-Enabled Target Detection}
Finally, we present the target detection in the active IRS-enabled sensing system with a point target located at the NLoS of the BS\cite{active_IRS_xianxin}. As an initial work on active IRS-enabled detection, the authors in \cite{active_IRS_xianxin}  assumed the target's angle $\theta$ and channel coefficient $\alpha$ are known and focused on determining the presence of the target at a given location. By collecting the receive signals $\mathbf y_3(t)$, IRS's reflection noise $\mathbf z(t)$, and IRS receive noise $\mathbf n_3(t)$ in the whole $T$ sensing symbols, we have $\tilde{\mathbf y}_3=[\mathbf y_3^T(1),\cdots,\mathbf y_3^T(T)]^T$, $\tilde{\mathbf z}=[\mathbf z^T(1),\cdots,\mathbf z^T(T)]^T$, and $\tilde{\mathbf n}_3=[\mathbf n_3^T(1),\cdots,\mathbf n_3^T(T)]^T$. The received echo signals at the IRS under hypotheses $\mathcal H_0$  and $\mathcal H_1$ are respectively given as
\begin{subequations}
  \begin{align}
&\mathcal H_0: \tilde{\mathbf y}_3=\tilde{\mathbf n}_3,\\\label{eq:echo_signal_piont_active}
&\mathcal H_1: \tilde{\mathbf y}_3= \left(\mathbf I_T \otimes\alpha\mathbf b(\theta)\mathbf a^{T}(\theta)\mathbf\Phi\mathbf G_t\right)\tilde{\mathbf x} + \left(\mathbf I_T \otimes\alpha\mathbf b(\theta)\mathbf a^{T}(\theta)\mathbf\Phi\right)\tilde{\mathbf z} + \tilde{\mathbf n}_3. 
  \end{align}
\end{subequations}
Then, the PDFs of the received echo signal under hypotheses $\mathcal H_0$ and $\mathcal H_1$ are respectively given as
\begin{small}
\begin{subequations}
  \begin{align}
&p(\tilde{\mathbf y}_3|\mathcal H_0)=\frac{\exp\left(-\frac{1}{\sigma^2}\|\tilde{\mathbf y}_3\|^2\right)}{\pi^{M_tT}\sigma^2},\\
&p(\tilde{\mathbf y}_3|\mathcal{H}_1) = \frac{\exp\!\left(\!-\!\left(\tilde{\mathbf y}_3\!-\!\left(\mathbf I_T\!\otimes\!\mathbf G_r\mathbf \Phi^T\alpha\mathbf b(\theta)\mathbf a^{T}(\theta)\mathbf\Phi\mathbf G_t\right)\tilde{\mathbf x}\right)^H\!(\mathbf C\!+\!\sigma^2\mathbf I_{M_rT})^{-1}\!\left(\tilde{\mathbf y}_3\!-\!\left(\mathbf I_T \!\otimes\!\mathbf G_r\mathbf \Phi^T\alpha\mathbf b(\theta)\mathbf a^{T}(\theta)\mathbf\Phi\mathbf G_t\right)\tilde{\mathbf x}\right)\!\right)}{\pi^{M_tT}\det(\mathbf C+\sigma^2\mathbf I_{M_rT})},
  \end{align}
\end{subequations}
\end{small}
where $\mathbf C = \mathbb{E}\left(\left(\left(\mathbf I_T \otimes\alpha\mathbf b(\theta)\mathbf a^{T}(\theta)\mathbf\Phi\right)\tilde{\mathbf z}\right)\left(\left(\mathbf I_T \otimes\alpha\mathbf b(\theta)\mathbf a^{T}(\theta)\mathbf\Phi\right)\tilde{\mathbf z}\right)^H\right)  =\mathbf I_{T}\otimes \sigma_z^2|\alpha|^2\|\mathbf a (\theta)\|^2\mathbf b^{T}(\theta)\mathbf\Phi\mathbf\Phi^H\mathbf b^{*}(\theta)$ denotes the covariance matrix of $\left(\mathbf I_T \otimes\alpha\mathbf b(\theta)\mathbf a^{T}(\theta)\mathbf\Phi\right)\tilde{\mathbf z}$. By applying the NP criterion, we decide hypothesis $\mathcal H_1$ if
\begin{equation}\label{eq:detector_3}
L(\tilde{\mathbf y}_3) = \frac{p(\tilde{\mathbf y}_3|\mathcal H_1)}{p(\tilde{\mathbf y}_3|\mathcal H_0)}> \delta_3.
\end{equation}
Otherwise, we decide hypothesis $\mathcal H_0$. With proper manipulations, \eqref{eq:detector_3} is equivalently expressed as \cite{active_IRS_xianxin}
\begin{equation}\label{eq:detector_3_re}
\tilde{\mathbf y}_3^H\frac{1}{\sigma^2}\mathbf C(\mathbf C+\sigma^2\mathbf I_{M_rT})^{-1}\tilde{\mathbf y}_3
+2\mathrm{Re}\left\{\left(\left(\mathbf I_T\!\otimes\!\mathbf G_r\mathbf \Phi^T\alpha\mathbf b(\theta)\mathbf a^{T}(\theta)\mathbf\Phi\mathbf G_t\right)\tilde{\mathbf x}\right)^H(\mathbf C+\sigma^2\mathbf I_{M_rT})^{-1}\tilde{\mathbf y}_3\right\}\stackrel{\mathcal{H}_1}{\underset{\mathcal{H}_0}{\gtrless}} \delta'_3,
\end{equation}
where $\delta'_3 = \ln\delta_3 + \left(\left(\mathbf I_T \otimes\mathbf G_r\mathbf \Phi^T\alpha\mathbf b(\theta)\mathbf a^{T}(\theta)\mathbf\Phi\mathbf G_t\right)\tilde{\mathbf x}\right)^H(\mathbf C+\sigma^2\mathbf I_{M_rT})^{-1}\left(\left(\mathbf I_T \otimes\mathbf G_r\mathbf \Phi^T\alpha\mathbf b(\theta)\mathbf a^{T}(\theta)\mathbf\Phi\mathbf G_t\right)\tilde{\mathbf x}\right)-M_rT\ln \sigma^{2}+\ln \det(\mathbf C+\sigma^2\mathbf I_{M_rT})$. Note that the left-hand-side in \eqref{eq:detector_3_re} consists of two components. The first component, i.e., $\tilde{\mathbf y}_3^H\frac{1}{\sigma^2}\mathbf C(\mathbf C+\sigma^2\mathbf I_{M_rT})^{-1}\tilde{\mathbf y}_3$, is an energy detector, corresponding to the utilization of the active IRS's reflection noise. The other component, i.e., $2\mathrm{Re}\left\{\left(\left(\mathbf I_T\!\otimes\!\mathbf G_r\mathbf \Phi^T\alpha\mathbf b(\theta)\mathbf a^{T}(\theta)\mathbf\Phi\mathbf G_t\right)\!\tilde{\mathbf x}\right)^H(\mathbf C\!+\!\sigma^2\mathbf I_{M_rT})^{-1}\tilde{\mathbf y}_3\right\}$, is a matched filter, corresponding to the utilization of the BS's transmit signal. The optimal detector design in the active IRS-enabled sensing system jointly utilizes the BS's transmit signal and the active IRS's reflection noise\cite{active_IRS_xianxin}. Based on the detector design in \eqref{eq:detector_3_re}, the false alarm and the detection probabilities are respectively given as
\begin{subequations}
  \begin{align}
 P_{3,\text{FA}} &= \mathcal{Q}_{\chi^2_{2T}(\lambda_1)}\left(\frac{2\delta'\left(1+\mathrm{tr}\left(\mathbf \Phi^H\mathbf H_2^H\mathbf H_2\mathbf\Phi\sigma_z^2/\sigma^2\right)\right)}{\mathrm{tr}\left(\mathbf \Phi^H\mathbf H_2^H\mathbf H_2\mathbf\Phi\sigma_z^2/\sigma^2\right)} + \lambda_1\right)\\
 P_{3,\text{D}} &= \mathcal{Q}_{\chi^2_{2T}(\lambda_2)}\left(\frac{2\delta'}{\mathrm{tr}\left(\mathbf \Phi^H\mathbf H_2^H\mathbf H_2\mathbf\Phi\right)\sigma_z^2/\sigma^2} + \frac{\lambda_1}{1+\mathrm{tr}\left(\mathbf \Phi^H\mathbf H_2^H\mathbf H_2\mathbf\Phi\right)\sigma_z^2/\sigma^2}\right)\\\label{eq:PD_active_FA}
&=\mathcal{Q}_{\chi^2_{2T}(\lambda_2)}\left(\frac{\mathcal{Q}^{-1}_{\chi^2_{2T}(\lambda_1)}\left(P_{3,\text{FA}}\right)}{1+\mathrm{tr}\left(\mathbf \Phi^H\mathbf H_2^H\mathbf H_2\mathbf\Phi\right)\sigma_z^2/\sigma^2}\right),
  \end{align}
\end{subequations}
where $\lambda_1=\frac{2T\text{SNR}_2^2\sigma^2}{\mathrm{tr}\left(\mathbf \Phi^H\mathbf H_2^H\mathbf H_2\mathbf\Phi\right)\sigma_z^2}$ and $\lambda_2=\lambda_1\left(1+\mathrm{tr}\left(\mathbf \Phi^H\mathbf H_2^H\mathbf H_2\mathbf\Phi\right)\sigma_z^2/\sigma^2\right)$\cite{active_IRS_xianxin}. Compared with the fully- and semi-passive IRS-enabled sensing systems, the signal amplification at the active reflecting elements generates additional noise to the transmit signal. In general, this reflection noise degrades the sensing SNR as shown in \eqref{eq:SNR_active}, which is harmful to sensing. However, for the specific target detection task, this additional noise can be utilized for detector design by measuring the energy of received signals. 
Therefore, with advanced detector design in \eqref{eq:detector_3_re} jointly exploiting the BS's transmit signal as well as the IRS's reflection noise, it is expected that active IRS achieves higher detection probability than fully- and semi-passive IRSs\cite{active_IRS_xianxin}. 

\subsection{Parameter Estimation}

In parameter estimation tasks, the objective is to extract valuable target's information by processing echo signals in \eqref{eq:echo_fully_passive}, \eqref{echo_semi_IRS}, and \eqref{echo_active_IRS}. For illustration, we consider the case with a point target at the NLoS region of the BS and aim to estimate the angle of target\cite{10008725,10138058,xianxin_comparsion}. The analysis framework is extendable to the case with other parameters to be estimated. For parameter estimation tasks, the estimation CRB is widely adopted as a performance metric to indicate the precision of estimation, establishing a lower bound on the variance of any unbiased estimators\cite{10217169,kay1993fundamentals,1703855,4359542,9652071,10251151,richards2014fundamentals}.

\subsubsection{Fully-Passive IRS-Enabled Target's Angle Estimation}
First, we present the parameter estimation problem with a point target located at the NLoS region of the BS\cite{10008725,10138058,xianxin_comparsion}. In this case, the target response matrix becomes $\mathbf H_1 =\alpha\mathbf a(\theta)\mathbf a^T(\theta)$. Let vector $\bm \eta =[\theta,\mathrm{Re}(\alpha),\mathrm{Im}(\alpha)]^T$ denote the unknown parameters of the point target. Typically, the estimation CRB is derived by first calculating the Fisher information matrix (FIM) for the unknown parameters and then obtaining the CRB by taking the inverse of the FIM\cite{kay1993fundamentals,richards2014fundamentals}. Let $\mathbf F \in \mathbb{R}^{3\times3}$ denote the FIM. Based the received signal model in \eqref{eq:echo_signal_piont_fully}, each element of $\mathbf F$ is 
\begin{equation}\label{eq:FIM}
\begin{split}
&\mathbf F_{i,k}=\mathrm{tr}\left(\mathbf R_n^{-1}\frac{\partial \mathbf R_n}{\partial \bm \eta_i}\mathbf R_n^{-1}\frac{\partial \mathbf R_n}{\partial \bm \eta_k}\right)\\
&+2\mathrm{Re}\left\{\frac{\partial \left(\left(\mathbf I_T \otimes\mathbf G_r\mathbf \Phi^T\alpha\mathbf a (\theta)\mathbf a^{T}(\theta)\mathbf\Phi\mathbf G_t\right)\tilde{\mathbf x} \right)^\mathrm{H}}{\partial \bm \eta_i}\mathbf R_n^{-1}\frac{\partial \left(\left(\mathbf I_T \otimes\mathbf G_r\mathbf \Phi^T\alpha\mathbf a (\theta)\mathbf a^{T}(\theta)\mathbf\Phi\mathbf G_t\right)\tilde{\mathbf x} \right)}{\partial \bm \eta_k}\right\}, i,k\in\{1,2,3\},
\end{split}
\end{equation} 
where $\mathbf R_n = \sigma^2\mathbf I_{M_rT}$ is the covariance matrix of the receive noise $\tilde{\mathbf n}_1$\cite{kay1993fundamentals,richards2014fundamentals,10008725,10138058}. Furthermore, we partition the FIM matrix $\mathbf F$ as
\begin{equation}\label{eq:FIM_partition}
\mathbf F=
\begin{bmatrix}
\mathbf{F}_{\theta \theta} & \mathbf{F}_{\theta \tilde{\bm\alpha}}\\
\mathbf{F}^{T}_{\theta \tilde{\bm\alpha}} & \mathbf{F}_{\tilde{\bm\alpha} \tilde{\bm\alpha}}
\end{bmatrix},
\end{equation}
where $\tilde{\bm\alpha} = [\mathrm{Re}(\alpha),\mathrm{Im}(\alpha)]^T$,
$\mathbf F_{\theta \theta}=\frac{2T|\alpha|^2}{\sigma^2}\mathrm{tr}\left(\dot {\mathbf B}_1(\theta) \mathbf R \dot {\mathbf B}^{H}_1(\theta)\right)$, $\mathbf F_{\theta \tilde{\bm \alpha}}=\frac{2T}{\sigma_\text{R}^2}\mathrm{Re}\{\alpha^*\mathrm{tr}\left(\mathbf B_1(\theta) \mathbf R  \dot {\mathbf B}^{H}_1(\theta)\right)[1,j]\}$, and $\mathbf F_{\tilde{\bm\alpha} \tilde{\bm\alpha}}=\frac{2T}{\sigma_\text{R}^2}\mathrm{tr}\left(\mathbf B_1(\theta) \mathbf R \mathbf B^{H}_1(\theta)\right)\mathbf I_2$. In \eqref{eq:FIM_partition}, $\mathbf B_1(\theta)= \mathbf p_r(\theta)\mathbf p_t^{T}(\theta)$, $\dot{\mathbf B}_1(\theta) =\dot{\mathbf p}_r(\theta)\mathbf p_t^{T}(\theta)+\mathbf p_r(\theta)\dot{\mathbf p}_t^{T}(\theta)$, $\mathbf p_t(\theta) = \mathbf G_t^{T} \mathbf \Phi^{T} \mathbf a(\theta)$, $\mathbf p_r(\theta) = \mathbf G_r \mathbf \Phi^{T} \mathbf a(\theta)$, $\dot{\mathbf p}_t(\theta) = \mathbf G_t^{T} \mathbf \Phi^{T} \dot{\mathbf a}(\theta)$, and $\dot{\mathbf p}_r(\theta) = \mathbf G_r \mathbf \Phi^{T} \dot{\mathbf a}(\theta)$ with $\dot{\mathbf a}(\theta)$ being the partial derivative of $\mathbf a(\theta)$ w.r.t. $\theta$\cite{10008725,10138058}. Based on the aforementioned FIM $\mathbf F$, the CRB for target's angle estimation is 
\begin{equation}\label{eq:FIM_theta}
\mathrm{CRB}_1(\theta) =[\mathbf F^{-1}]_{1,1} =[\mathbf F_{\theta \theta}-\mathbf F_{\theta \tilde{\bm\alpha}}\mathbf F_{\tilde{\bm\alpha} \tilde{\bm\alpha}}^{-1}\mathbf F_{\theta \tilde{\bm\alpha}}^\mathrm{T}]^{-1}.
\end{equation}
Through proper calculations, CRB for target's angle estimation in the fully-passive IRS-enabled sensing system is \cite{10008725,10138058,xianxin_comparsion}
\begin{subequations}
  \begin{align}
\mathrm{CRB}_1(\theta)&=\frac{\sigma^2}{2T|\alpha|^2\left(\mathrm{tr}\left(\dot {\mathbf B}_1(\theta) \mathbf R \dot {\mathbf B}^{H}_1(\theta)\right)-\frac{\left|\mathrm{tr}\left(\mathbf B_1(\theta) \mathbf R  \dot {\mathbf B}^{H}_1(\theta)\right)\right|^2}{\mathrm{tr}\left(\mathbf B_1(\theta) \mathbf R \mathbf B^{H}_1(\theta)\right)}\right)}\\\label{eq:CRB_fully-passive}
&=\frac{\sigma^2/(2T|\alpha|^2)}{\mathbf p_t^{H}(\theta)\mathbf R^{T}\mathbf p_t(\theta)\left(\|\dot{\mathbf p}_r(\theta)\|^2-\frac{|\dot{\mathbf p}_r^{H}(\theta)\mathbf p_r(\theta)|^2}{\|\mathbf p_r(\theta)\|^2}\right)+\|\mathbf p_r(\theta)\|^2\left(\dot{\mathbf p}_t^{H}(\theta) \mathbf R^{T}\dot{\mathbf p}_t(\theta) -\frac{|\mathbf p_t^{H}(\theta) \mathbf R^{T}\dot{\mathbf p}_t(\theta) |^2}{\mathbf p_t^{H}(\theta) \mathbf R^{T}\mathbf p_t(\theta) }\right)},
\end{align}
\end{subequations}
which decreases inversely proportionally to $N^6$ when the number of reflecting elements $N$ is sufficiently large\cite{xianxin_comparsion}. 

\subsubsection{Semi-Passive IRS-Enabled Target's Angle Estimation}
Next, we introduce the parameter estimation problem with a point target located at the NLoS region of the BS\cite{xianxin_comparsion}. In this case, the target response matrix is becomes $\mathbf H_2 =\alpha\mathbf a(\theta)\mathbf b^T(\theta)$. Similar for the CRB derivation with the fully-passive IRS, the CRB for target's angle estimation in the semi-passive IRS-enabled sensing system is
\begin{subequations}
  \begin{align}
\mathrm{CRB}_2(\theta)&=\frac{\sigma^2}{2T|\alpha|^2\left(\mathrm{tr}\left(\dot {\mathbf B}_2(\theta) \mathbf R \dot {\mathbf B}^{H}_2(\theta)\right)-\frac{\left|\mathrm{tr}\left(\mathbf B_2(\theta) \mathbf R  \dot {\mathbf B}^{H}_2(\theta)\right)\right|^2}{\mathrm{tr}\left(\mathbf B_2(\theta) \mathbf R \mathbf B^{H}_2(\theta)\right)}\right)}\\
\label{eq:CRB_semi-passive}
&=\frac{\sigma^2/(2T|\alpha|^2)}{\mathbf p_t^{H}(\theta) \mathbf R^{T}\mathbf p_t (\theta)\left(\|\dot{\mathbf b}(\theta)\|^2-\frac{|\dot{\mathbf b}^{H}(\theta)\mathbf b(\theta)|^2}{\|\mathbf b(\theta)\|^2}\right)+\|\mathbf b(\theta)\|^2\left(\dot{\mathbf p}_t^{H}(\theta) \mathbf R^{T}\dot{\mathbf p}_t(\theta) -\frac{|\mathbf p_t^{H}(\theta) \mathbf R^{T}\dot{\mathbf p}_t(\theta) |^2}{\mathbf p_t^{H}(\theta) \mathbf R^{T}\mathbf p_t(\theta) }\right)},
\end{align}
\end{subequations}
where  $\mathbf B_2(\theta)= \mathbf b(\theta)\mathbf p_t^{T}(\theta)$ and $\dot{\mathbf B}_2(\theta) =\dot{\mathbf b}(\theta)\mathbf p_t^{T}(\theta)+\mathbf b(\theta)\dot{\mathbf p}_t^{T}(\theta)$ with $\dot{\mathbf b}(\theta)$ being the partial derivative of $\mathbf b(\theta)$ w.r.t. $\theta$\cite{xianxin_comparsion}. By comparing the differences of CRBs in \eqref{eq:CRB_fully-passive} and \eqref{eq:CRB_semi-passive}, i.e.,  $\|\mathbf p_r(\theta)\|^2$ versus $\|\mathbf b(\theta)\|^2$,  $\|\dot{\mathbf p}_r(\theta)\|^2$ versus $\|\dot{\mathbf b}(\theta)\|^2$, and $|\dot{\mathbf p}_r^{H}(\theta)\mathbf p_r(\theta)|^2$ versus $|\dot{\mathbf b}^{H}(\theta)\mathbf b(\theta)|^2$, fully-passive IRS is expected to benefits from more reflective beamforming than semi-passive IRS.
Interesting, it has been shown in \cite{xianxin_comparsion} that the $\mathrm{CRB}_2(\theta)$ decreases inversely proportionally to $N^4$ when the number of reflecting elements is sufficiently large, which is less than the beamforming gain with fully-passive IRS. This is due to the fact that semi-passive IRS-enabled sensing system lacks the reflective beamforming gain from the IRS to the BS.

\subsubsection{Active IRS-Enabled Target's Angle Estimation}
Finally, for the active IRS-enabled sensing systems with a point target located at the NLoS region of the BS, the target response matrix is modeled as $\mathbf H_2 =\alpha\mathbf a(\theta)\mathbf b^T(\theta)$. Based on the received echo signal model in \eqref{eq:echo_signal_piont_active}, when the power of IRS' reflection noise $\sigma_z^2$ approaches the power of noise at receiver $\sigma^2$, we can omit the term $\mathbf H\mathbf\Phi\mathbf z(t)$ in \eqref{echo_active_IRS} considering the huge path loss of the IRS-target-IRS link\cite{10497119,10496515,fang2024joint}. Under this approximation, the received echo signal at the IRS is approximated as 
\begin{equation}
\begin{split}
\tilde{\mathbf y}_3 &= \left(\mathbf I_T \otimes\alpha\mathbf b(\theta)\mathbf a^{T}(\theta)\mathbf\Phi\mathbf G_t\right)\tilde{\mathbf x} + \left(\mathbf I_T \otimes\alpha\mathbf b(\theta)\mathbf a^{T}(\theta)\mathbf\Phi\right)\tilde{\mathbf z} + \tilde{\mathbf n}_3\\
&\approx \left(\mathbf I_T \otimes\alpha\mathbf b(\theta)\mathbf a^{T}(\theta)\mathbf\Phi\mathbf G_t\right)\tilde{\mathbf x}  + \tilde{\mathbf n}_3. 
\end{split}
\end{equation}
With the aforementioned approximation, the CRB for target's angle estimation in active IRS-enabled sensing system is approximated as 
\begin{subequations}
  \begin{align}
\mathrm{CRB}_3(\theta)&\approx\frac{\sigma^2}{2T|\alpha|^2\left(\mathrm{tr}\left(\dot {\mathbf B}_2(\theta) \mathbf R \dot {\mathbf B}^{H}_2(\theta)\right)-\frac{\left|\mathrm{tr}\left(\mathbf B_2(\theta) \mathbf R  \dot {\mathbf B}^{H}_2(\theta)\right)\right|^2}{\mathrm{tr}\left(\mathbf B_2(\theta) \mathbf R \mathbf B^{H}_2(\theta)\right)}\right)}\\
\label{eq:CRB_active-passive}
&=\frac{\sigma^2/(2T|\alpha|^2)}{\mathbf p_t^{H}(\theta) \mathbf R^{T}\mathbf p_t (\theta)\left(\|\dot{\mathbf b}(\theta)\|^2-\frac{|\dot{\mathbf b}^{H}(\theta)\mathbf b(\theta)|^2}{\|\mathbf b(\theta)\|^2}\right)+\|\mathbf b(\theta)\|^2\left(\dot{\mathbf p}_t^{H}(\theta) \mathbf R^{T}\dot{\mathbf p}_t(\theta) -\frac{|\mathbf p_t^{H}(\theta) \mathbf R^{T}\dot{\mathbf p}_t(\theta) |^2}{\mathbf p_t^{H}(\theta) \mathbf R^{T}\mathbf p_t(\theta) }\right)}.
\end{align}
\end{subequations}
The structure of CRB for target's angle estimation in active IRS-enabled sensing system, $\mathrm{CRB}_3(\theta)$, is similar as that in semi-passive IRS-enabled sensing system, $\mathrm{CRB}_2(\theta)$. It is expected that $\mathrm{CRB}_3(\theta)$ decreases inversely proportionally to $N^4$ when the number of reflecting elements $N$ is sufficiently large, similar as that in semi-passive IRS-enabled sensing system.

\subsection{Joint Transmit and Reflective Beamforming Design}
Given the above closed-form expressions of sensing performance related to the BS's transmit signal and IRS's reflection, jointly optimizing the transmit and reflective beamforming can enhance the sensing performance, which, however, is a challenging task. For target detection and parameter estimation tasks, our objective is to maximize the detection probability and  minimize the estimation CRB, respectively, by joint beamforming design, subject to the constraints on the BS's transmit signal and the IRS's reflection.

In order to facilitate the joint transmit beamforming design at the BS and reflective beamforming design at the IRS, the channel channel state information (CSI) of transmitter-IRS and IRS-receiver links should be efficient estimated. Some prior works \cite{8937491,9195133,10465210,9722893,10443321} have studied the CSI estimation problem using different IRS architectures. First, with fully-passive IRS, the cascaded transmitter-IRS-receiver channel can be estimated by changing the reflection pattern of the IRS with pilot signals, measuring the received signal at the receivers, and recovering the cascaded channel information from the received signal\cite{8937491,9195133,10465210}. Furthermore, in order to estimate the separate transmitter-IRS and IRS-receiver channels efficiently, some priors works \cite{9722893,10443321} proposed semi-passive IRS architecture by deploying dedicated sensors at the IRS in addition to the reflecting elements. Then, the separate channel from the transmitter/receivers to the IRS can be estimated by processing the received signals at the IRS's sensors.

With the CSI, the joint transmit and reflecting beamforming design problem is still challenging due to the non-convex nature of the problem, and the coupled relation between transmit and reflecting beamforming design. First, the transmit beamforming and the reflective beamforming are strongly coupled with each other. It is challenging to decouple the transmit and reflective beamforming optimization problem. As a compromise, alternating optimization (AO) methods are wildly used to optimize the transmit and reflective beamforming iteratively, with the other one given in each iteration\cite{8811733,9427474,10159991,10008725,10138058,10464564,xianxin_comparsion,9771801,10279464,10440056,10497119,9416177,9364358,10254508,9769997,10086570,9591331,10050406,9979782,10054402,fang2024joint,10496515,10319318,10184278,10186271,10149664,10304580
}. However, AO methods is only capable of obtaining a suboptimal solution, and its performance gap to the global optimal solution is still unknown in literature. Second, the target detection probability or estimation CRB generally has an implicit interplay with system parameters. Then, the intricate performance metric expressions together with several practical considerations on the reflection of IRS usually result in non-convex constraints. Various non-convex optimization methods, such as successive convex approximation (SCA) methods\cite{6675875,razaviyayn2014parallel,razaviyayn2014successive,8752072}, can be exploited to address these highly non-convex optimization problems\cite{10008725,10138058,10464564,xianxin_comparsion,10279464,10440056,10497119,fang2024joint}. In SCA methods, the non-convex optimization problem is first approximated by a series of convex optimization problems, which are then iteratively solved before converging to a suboptimal solution. However, these non-convex optimization methods normally are of high computational complexity and their performance depends heavily on the initial point chosen. How to obtain an efficient joint beamforming design with reduced computational complexity by exploiting a learning-based approach is worthy of future research.

\subsubsection{Joint Beamforming Design for Detection Probability Maximization} \label{subsec:joint_BF_sensing}
To provide more insights, we present the joint beamforming designs for detection probability maximization with a point target located at the NLoS region at the BS, when the channel between the BS and the IRS is LoS\cite{xianxin_comparsion,active_IRS_xianxin}. In this case, the channel matrices $\mathbf G_t$ and $\mathbf G_r$ are modeled as $\mathbf G_t = \sqrt{L(d_1)} \mathbf c \mathbf d^T$ and $\mathbf G_r = \sqrt{L(d_1)} \mathbf e \mathbf c^T$, where vector $\mathbf c\in\mathbb C^{N\times 1}$ denotes the steering vector of IRS reflecting elements towards the BS, vector $\mathbf d\in\mathbb C^{N\times 1}$ denotes the steering vector of transmit antennas at the BS towards the IRS, and vector $\mathbf e\in\mathbb C^{N\times 1}$ denotes the steering vector of receive antennas at the BS towards the IRS.

First, for the fully-passive IRS-enabled sensing system, maximizing detection probability is equivalent to maximizing sensing SNR. Based on the LoS channel models and the target response matrix of point target, the sensing SNR with fully-passive is expressed as
\begin{equation}\label{eq:SNR_LoS_fully}
\text{SNR}_1 = 
\frac{\mathrm{tr}\left(\mathbf G_r\mathbf \Phi^T\mathbf H_1\mathbf\Phi\mathbf G_t\mathbf R\mathbf G_t^H\mathbf \Phi^H\mathbf H_1^H\mathbf \Phi^*\mathbf G_r^H\right)}{\sigma^2}=\frac{|\alpha|^2L^2(d_1)\|\mathbf e\|^2|\mathbf a^T(\theta)\mathbf\Phi\mathbf c|^4\mathbf d^T\mathbf R\mathbf d^*}{\sigma^2}.
\end{equation}
We aim to maximize the sensing SNR by jointly optimizing the transmit beamforming at the BS, i.e., $\mathbf R$, and the reflective beamforming at the IRS, i.e., $\mathbf \Phi$. In this specific channel model, the transmit and reflective beamforming design can be decoupled into two subproblems, i.e., the maximizations of $\mathbf d^T\mathbf R\mathbf d^*$ and $|\mathbf c^T\mathbf\Phi\mathbf a(\theta)|^4$. In this case, the optimal transmit beamforming design is maximum ratio transmission (MRT), i.e., $\mathbf R_\text{opt}=\frac{P_\text{BS}\mathbf d^*\mathbf d^{T}}{\|\mathbf d\|^2}$, and the optimal reflective beamforming design aligns the phase of signals arriving the target, i.e., $\mathbf \Phi_\text{opt}= e^{-j\mathrm{arg}(\mathbf c+\mathbf a(\theta))}$\cite{xianxin_comparsion}. By substituting the optimal joint beamforming design back into the sensing SNR in \eqref{eq:SNR_LoS_fully}, we have
\begin{equation}
\text{SNR}_1^\text{opt} =\frac{|\alpha|^2L^2(d_1)M_tM_rN^4}{\sigma^2},
\end{equation}
which increases proportionally to $N^4$\cite{xianxin_comparsion}.

Next, we introduce the joint beamforming design for detection probability maximization in semi-passive IRS-enabled sensing system. Under the aforementioned LoS channel and point target models, the sensing SNR in \eqref{eq:SNR_semi} becomes
\begin{equation}\label{eq:SNR_LoS_semi}
\text{SNR}_2 =\frac{\mathrm{tr}\left(\mathbf H_2\mathbf\Phi\mathbf G_t\mathbf R\mathbf G_t^H\mathbf \Phi^H\mathbf H_2^H\right)}{\sigma^2}=\frac{|\alpha|^2L(d_1)\|\mathbf b(\theta)\|^2|\mathbf a^T(\theta)\mathbf\Phi\mathbf c|^2\mathbf d^T\mathbf R\mathbf d^*}{\sigma^2}.
\end{equation}
Based on the sensing SNR in \eqref{eq:SNR_LoS_semi}, the joint beamforming designs for SNR maximization or equivalently for detection probability maximization are $\mathbf R_\text{opt}=\frac{P_\text{BS}\mathbf d^*\mathbf d^{T}}{\|\mathbf d\|^2}$ and $\mathbf \Phi_\text{opt}= e^{-j\mathrm{arg}(\mathbf c+\mathbf a(\theta))}$\cite{xianxin_comparsion}. In this case, the maximum sensing SNR with semi-passive IRS is 
\begin{equation}
\text{SNR}_2^\text{opt} =\frac{|\alpha|^2L(d_1)M_tM_rN^2}{\sigma^2},
\end{equation}
which increases proportionally to $N^2$\cite{xianxin_comparsion}.

Finally, we introduce the joint beamforming design for detection probability maximization with active IRS\cite{active_IRS_xianxin}. To maximize the detection probability in \eqref{eq:PD_active_FA}, the optimal transmit beamforming is $\mathbf R_\text{opt}=\frac{P_x\mathbf d^*\mathbf d^{T}}{\|\mathbf d\|^2}$ with $P_x =\min\big\{P_\text{BS}, (P_\text{IRS}/Na_0^2-\sigma_z^2)/(L(d_1)M_t)\big\}$ and the optimal reflective beamforming is $\mathbf \Phi_\text{opt}= a_0 e^{-j\mathrm{arg}(\mathbf c+\mathbf a(\theta))}$, where $a_0$ is the amplification gain at each reflecting element and its values can be obtained by one-dimensional search over the feasible region to maximize the detection probability\cite{active_IRS_xianxin}. It is observed that the optimal transmit beamforming follows the MRT structure with the power constraints at the BS and the active IRS. Meanwhile, the optimal reflective beamforming aligns the phase of signals pointing to the target, with the same amplification gain at each reflecting element.

\subsubsection{Joint Beamforming Design for Estimation CRB Minimization}\label{Joint_BF_CRB_min}
In this subsection, we introduce the joint beamforming design for CRB minimization briefly. Due to the complex expressions of CRB, it is challenging to decouple the transmit and reflective beamforming design problems. To solve this issue, the AO method is employed to alternately optimize the transmit and reflective beamforming\cite{9771801,10138058,xianxin_comparsion}. In particular, with any given reflective beamforming design, the transmit beamforming optimization problem is convex and can be optimally solved using CVX\cite{cvx}. However, the reflective beamforming optimization problem is highly non-convex with any given transmit beamforming design, in which a converged solution is obtained using semi-definite relaxation (SDR)\cite{5447068} and SCA technologies\cite{9771801,10138058,xianxin_comparsion}.

\subsection{Performance Evaluation}  
This subsection evaluates the target detection and angle estimation performance with fully-passive, semi-passive, and active IRSs, for gaining insights.

\begin{figure}[ht]
        \centering
        \includegraphics[width=0.5\textwidth]{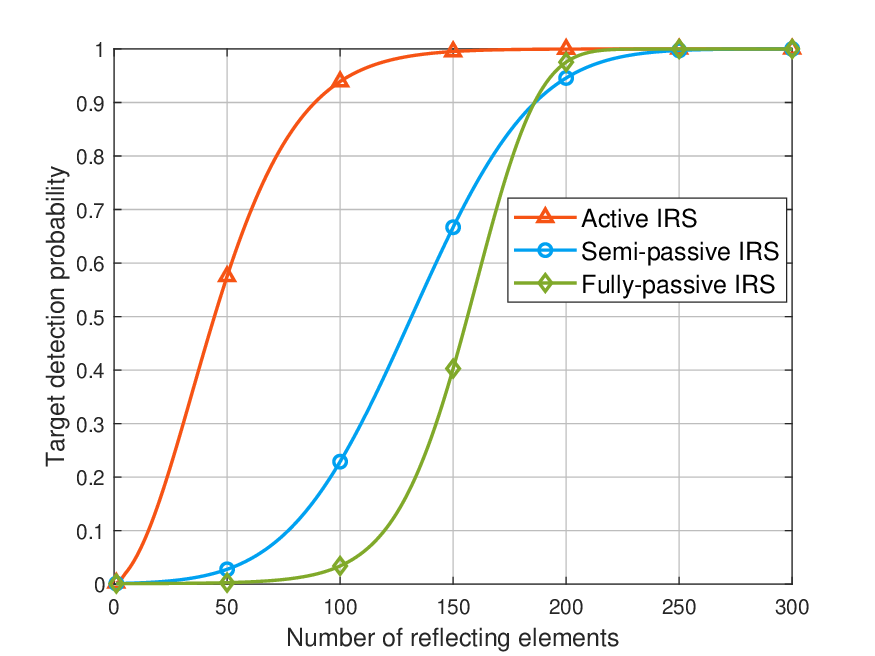}
        \caption{The target detection probability versus the number of reflecting elements $N$ equipped at the IRS.}
        \label{PD_elements}
\end{figure}

First, we present the detection performance performance with joint beamforming optimization for an IRS-enabled NLoS sensing system consisting of a multi-antenna BS, an IRS, and a point target located at the NLoS region of the BS. We assume that the BS-IRS and IRS-BS links are LoS. The maximum BS's transmit power and IRS's reflection power with active IRS are set as $P_\text{BS} = 1$~W\cite{10138058,10287779,10443321,xianxin_comparsion} and $P_\text{IRS}=0.1$~W\cite{9979782,10319318,9896755}, respectively. For a fair comparison, for the cases with fully- and semi-passive IRSs, the maximum transmit power at the BS is set as $P_\text{BS} = 1.1$~W. The path loss at distance $d$ is modeled as $L(d)=K_0\left(\frac{d}{d_0}\right)^{-\alpha_0}$, where $K_0=-30~\text{dB}$ denote the path loss at the reference distance $d_0=1~\text{m}$ \cite{8811733,9365004,10138058,9427474,xianxin_comparsion}. The path loss exponent of the BS-IRS and IRS-target links is $\alpha_0=2.2$\cite{9979782,xianxin_comparsion,10496515,9896755}. The distance of the BS-IRS and IRS-target links are set as $d_1 = 5$~m and $d_2=65$~m, respectively. We also set $M_t=M_r=8$, $T=128$, $\sigma^2=-80$~dBm\cite{8811733,9365004,10496515}, and $\sigma_z^2=-30$~dBm\cite{active_IRS_xianxin}, respectively. With the aforementioned simulation setup and joint beamforming design for detection probability maximization in Section~\ref{subsec:joint_BF_sensing}, the detection probabilities with fully-passive, semi-passive, and active IRSs are given in Figure~\ref{PD_elements}.
It is observed that the detection probabilities of all three types of IRS increase with the number of reflecting elements $N$. Meanwhile, due to the signal power amplification provided by the active IRS and the advanced detector design \cite{active_IRS_xianxin} jointly utilizing the BS's transmitted signal and the reflection noise, the detection performance of active IRS outperforms the other two types at the IRS. In addition, the detection performance with fully-passive IRS is observed to exceed that of the semi-passive counterpart when $N$ is sufficiently large.

\begin{figure}[ht]
        \centering
        \includegraphics[width=0.5\textwidth]{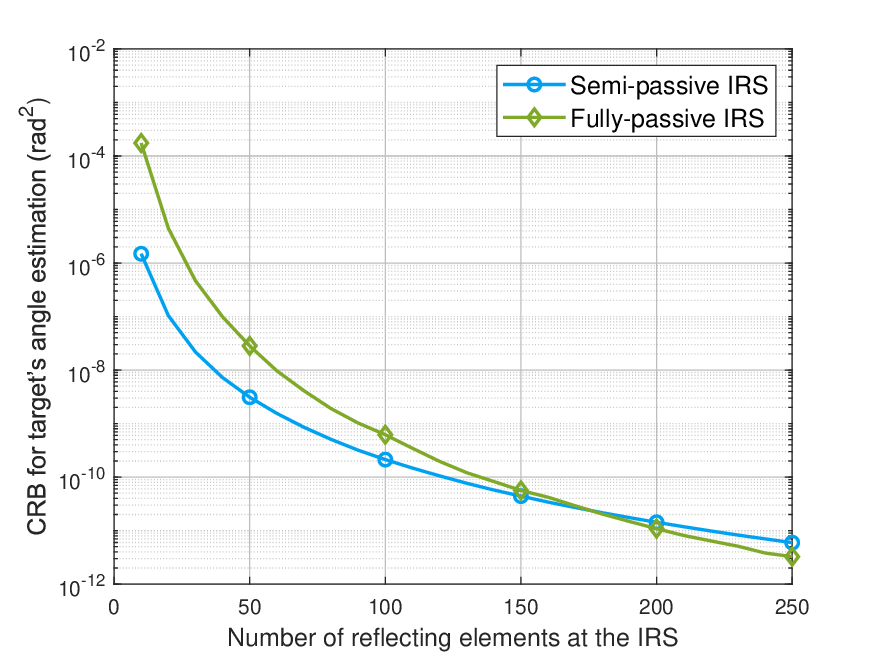}
        \caption{The CRB for target's angle estimation versus the number of reflecting elements $N$ equipped at the IRS.}
        \label{CRB_elements}
\end{figure}

Then, we provide the target's angle estimation performance with joint beamforming optimization for an IRS-enabled NLoS sensing system consisting of a multi-antenna BS, an IRS, and a point target located at the NLoS region of the BS\cite{xianxin_comparsion}. For the BS-IRS and IRS-target links, the path loss exponent $\alpha_0$ is set as $2.2$\cite{9979782,xianxin_comparsion,10496515,9896755} and $2.0$\cite{8811733,xianxin_comparsion}, respectively. The distance of the BS-IRS and IRS-target links are set as $d_1 = \sqrt{2}$~m and $d_2=6$~m, respectively. We also set $M_t=M_r=4$, $T=256$, $P_\text{BS} =30~\text{dBm}$\cite{10138058,10287779,10443321,xianxin_comparsion}, and $\sigma^2 = -90~\text{dBm}$\cite{xianxin_comparsion}, respectively. With the aforementioned simulation setup and the joint beamforming designs in Section~\ref{Joint_BF_CRB_min}, CRBs for target's angle estimation with fully- and semi-passive IRSs are displayed in Figure~\ref{CRB_elements}. It is demonstrated that the estimation CRB with a fully-passive IRS is lower than that with a semi-passive IRS when the number of reflecting elements exceeds a certain threshold. This improvement is attributed to the enhanced reflective beamforming capabilities of the fully-passive IRS\cite{xianxin_comparsion}.

\section{IRS-Enabled ISAC}
\begin{figure}[ht]
        \centering
        \includegraphics[width=1.0\textwidth]{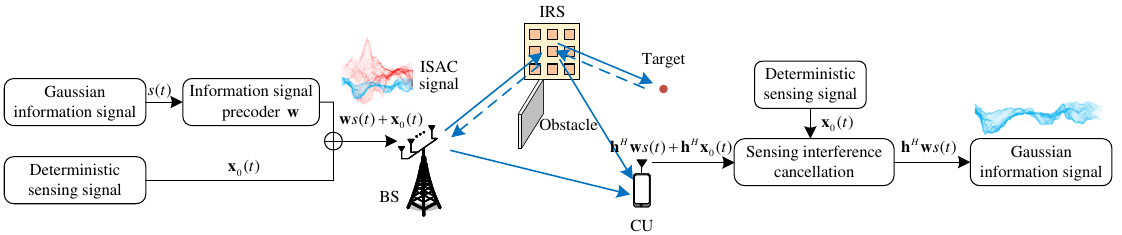}
        \caption{A basic architecture of IRS-enabled ISAC.}
        \label{IRS_enabled_ISAC_architectures}
\end{figure}

Considering the great benefits of IRS for S\&C systems, the IRS-enabled ISAC has also attracted a growing research attention recently\cite{9771801,10077119,10243495,10702570,10279464,10440056,10497119,10197455,9416177,9364358,10254508,9769997,9729741,9591331,10086570,10050406,9979782,10054402,10319318,10496515,10184278,10186271,10149664,10226306,10304580}. In particular, a basic IRS-enabled ISAC system architecture is shown in Figure~\ref{IRS_enabled_ISAC_architectures}, in which an IRS is deployed at an appropriate location to assist wireless communication and NLoS target sensing simultaneously. Besides the benefits of IRS for sensing, it also brings some benefits for ISAC. First, an extra transmission link between transceivers is established with the reflection of IRS, which extends the communication regions, improves the channel rank between transceivers, and increases the power of received signals. Second, the strategic reflective beamforming design at IRS can mitigate the interference between S\&C by manipulating the S\&C channels from strong correlation to weak correlation, and also enhance the coupling of S\&C channels, thereby improving the ISAC performance.

\subsection{Joint Transmit and Receive Design for ISAC}
For S\&C systems, the transmit signal design criteria are generally different. For communication systems, the BSs aim to transmit unknown messages to the CUs as reliably as possible, such that information-theoretic metrics such as the channel capacity or Shannon capacity are important for assessing the fundamental performance\cite{6773024,goldsmith2005wireless}. For the typical additive white Gaussian noise (AWGN) wireless communication channel, to achieve the Shannon capacity, the transmitted information signals follow the Gaussian distribution\cite{goldsmith2005wireless}. By contrast, for radar sensing, the BSs intend to obtain precise knowledge about the environment, such as whether a target is present in the interested region or the parameters of the target\cite{kay1993fundamentals,richards2014fundamentals,kay1993fundamentalsdetection}. Normally, detection probability and estimation CRB are adopted as performance metrics\cite{kay1993fundamentals,richards2014fundamentals,kay1993fundamentalsdetection}. For the target sensing tasks, to enhance the sensing performance, the BSs need to transmit deterministic sensing signals, which is different from communication systems tending to employ random (e.g., Gaussian)  signals for information transmission\cite{kay1993fundamentals,richards2014fundamentals,kay1993fundamentalsdetection}. Due to the different types of signals required for S\&C, only exploiting the traditional information signal or sensing signal for ISAC cannot achieve the ISAC performance bound. Therefore, to balance S\&C performance and achieve the ISAC performance bound, a viable solution is to allow the BS to transmit joint signals consisting of both information signals and dedicated sensing signals. 

A typical transmit and receive design for ISAC is given in Figure~\ref{IRS_enabled_ISAC_architectures}, in which the BS transmits joint signals for information transmission and target sensing. Let $s(t)$ denote the transmit information signal at symbol $t\in\mathcal{T}$, which follows Gaussian distribution with zero mean and unit variance for capacity achieving. Accordingly, let $\mathbf w \in \mathbb C^{M_t\times 1}$ denote the precoder at the BS for the information signal. Besides, let $\mathbf x_0(t)\in \mathbb C^{M_t\times 1}$ denote the dedicated sensing signal at symbol $t\in\mathcal{T}$, which is typically deterministic to achieve higher sensing performance. The sample covariance matrix of sensing signals is
\begin{equation}
\mathbf R_0=\frac{1}{T}\sum_{t=1}^T\mathbf x_0(t)\mathbf x_0^H(t).
\end{equation}  Then, the transmit joint signal at the BS is  modeled as
\begin{equation}
\mathbf x(t)=\mathbf ws(t)+\mathbf x_0(t),~t \in \mathcal T.
\end{equation}
The channel vectors from the BS and the IRS to the single antenna CU are respectively denoted as $\mathbf h_\text{d}\in \mathbb C^{M_t\times 1}$ and $\mathbf h_\text{r}\in \mathbb C^{N\times 1}$. Then, the combined channel from the BS to the CU is 
\begin{equation}
\mathbf h^H=\mathbf h_\text{d}^H+\mathbf h_\text{r}^H\mathbf\Phi\mathbf G_t.
\end{equation}
Thus, the received signal at the CU is 
\begin{equation}\label{eq:interference_cu}
y_c(t)=\left(\mathbf h_\text{d}^H+\mathbf h_\text{r}^H\mathbf\Phi\mathbf G_t\right)\mathbf ws(t)+\left(\mathbf h_\text{d}^H+\mathbf h_\text{r}^H\mathbf\Phi\mathbf G_t\right)\mathbf x_0(t)+ n_c(t), ~t \in \mathcal T,
\end{equation} 
where $n_c(t)$ is the noise at the CU's receiver with $\sigma_c^2$ being the power at each antenna. Note that, the received signal at the CU given in \eqref{eq:interference_cu} suffers from the interference from the dedicated sensing signal $\mathbf x_0(t)$. Normally, the dedicated sensing signal $\mathbf x_0(t)$ is deterministic, thus, it can be pre-known by the CU if the sensing data is sent to the CU before the ISAC transmission. As shown in Figure~\ref{IRS_enabled_ISAC_architectures}, with perfect CSI at the receiver, the CU can cancel the sensing signal interference by subtracting the component $\left(\mathbf h_\text{d}^H+\mathbf h_\text{r}^H\mathbf\Phi\mathbf G_t\right)\mathbf x_0(t)$ from the received signal $y_c(t)$\cite{cheng2024networked,10086626,9771801,10279464,10440056,10497119}. Based on this, we present two types of CU receiver, namely type-I and type-II CU receiver, which has not or has the ability to cancel the interference from the dedicated sensing signal, respectively\cite{cheng2024networked,10086626,9771801,10279464,10440056,10497119}. The SNRs at these two types of CU receiver are respectively expressed as follows
\begin{subequations}
  \begin{align}
  \gamma_{\text{I}}&=\frac{\left|\left(\mathbf h_\text{d}^H+\mathbf h_\text{r}^H\mathbf\Phi\mathbf G_t \right)\mathbf w\right|^2}{\left(\mathbf h_\text{d}^H+\mathbf h_\text{r}^H\mathbf\Phi\mathbf G_t \right)\mathbf R_0 \left(\mathbf h_\text{d}+\mathbf G_t^H\mathbf\Phi^H\mathbf h_\text{r} \right) +\sigma_c^2},\\
\gamma_{\text{II}}&= \frac{\left|\left(\mathbf h_\text{d}^H+\mathbf h_\text{r}^H\mathbf\Phi\mathbf G_t \right)\mathbf w\right|^2}{\sigma_c^2}.
 \end{align}
 \end{subequations}
Due to sensing interference cancellation ability at the CU,  type-II CU receiver can enhance the communication rate significantly\cite{cheng2024networked,10086626,9771801,10279464,10440056,10497119}. 

While for target sensing, under such joint signal design with information and dedicated sensing signal compensates, both communication signals and dedicated sensing signals can be utilized to enhance the sensing with advanced sensing algorithm design\cite{9124713,cheng2024networked,10086626,10153696,9916163,hua2024near,9771801,10279464,10440056,10497119,10254508,10050406,9979782,10496515
}. The sample covariance matrix of the transmitted joint signal over the $T$ symbols is 
\begin{equation}\label{eq:sample_covariance_matrix_ISAC}
\mathbf R = \frac{1}{T} \sum_{t=1}^T \left(\mathbf ws(t)+\mathbf x_0(t)\right)\left(\mathbf ws(t)+\mathbf x_0(t)\right)^H \approx \mathbf w \mathbf w^H + \mathbf R_0, 
\end{equation}
where the approximation is valid when the length of symbols $T$ is sufficiently large\cite{9652071,10596930}. 

\subsection{Fundamental Sensing versus Communication Performance Tradeoff}
As the ISAC system integrates both S\&C functionalities into a system, it is necessary to balance the S\&C performance metrics, and explore the ISAC performance boundaries. For notational convenience, let $C_\text{com}\left(\mathbf w, \mathbf R_0,\mathbf \Phi\right)$ and $C_\text{sen}\left(\mathbf w, \mathbf R_0,\mathbf \Phi\right)$ represent the communication and sensing performance indicators, respectively. For communication systems, the SNR at the CU receiver is a typical performance metric. For wireless sensing systems, target detection probability and estimation CRB are typical sensing performance as given in Section~\ref{sec:sensing}. Let $\mathcal C$ denote the sensing-communication performance region of IRS-enabled ISAC system, which is the combination of achieved sensing and communication performances under given transmit signal and IRS's reflection design. The sensing-communication performance region $\mathcal C$ is defined as
\begin{equation}
\mathcal C \triangleq \left\{(c_\text{com},c_\text{sen})|c_\text{com}\le C_\text{com}\left(\mathbf w, \mathbf R_0,\mathbf \Phi\right),c_\text{sen}\le C_\text{sen}\left(\mathbf w, \mathbf R_0,\mathbf \Phi\right),\left\{\mathbf w, \mathbf R_0\right\}\in \mathcal R_1,\mathbf \Phi\in \mathcal R_2\right\},
\end{equation}
where $\mathcal R_1$ and $\mathcal R_2$ denote the feasible regions of transmit and reflective beamforming designs, respectively. To obtain the Pareto boundaries \cite{4558045,9365004,10159991,10217169} for IRS-enabled ISAC systems, we aim to jointly optimize the transmit and reflective beamforming to maximize the communication SNR, subject to the sensing performance constraints and the feasible region of transmit signal and IRS's reflection. The sensing performance-constrained communication performance optimization problem is usually non-convex due to the non-convex constraint for the IRS's reflecting matrix and the non-convex S\&C performance metrics. Thus, how to explore the Pareto boundaries of ISAC performance is challenging.

\subsection{Joint Transmit and Reflective Beamforming Design}
Compared with the IRS-enabled sensing system, the joint transmit and reflective beamforming design problem in ISAC is more complicated. First, the presence of both information and sensing signals complicates the transmit signal design. Typically, to provide more degrees of freedoms (DoFs) for target sensing, there  are no restrictions on the rank of the sample covariance matrix for the dedicated sensing signal component, i.e., $\mathrm{rank}(\mathbf R_0)$. Normally, the transmit beamforming optimization problem with any given reflecting beamforming in IRS-enabled sensing is an semi-definite program (SDP)\cite{8811733,9427474,10008725,10138058,10464564,fang2024joint,9771801,10440056,10497119,10279464,9416177,10254508,10050406,10319318}. This problem is convex and can be optimally solved. While for wireless communications, the sample covariance matrix for the communication signal component should satisfy the rank-one constraint, i.e., $\mathrm{rank}(\mathbf w\mathbf w^H) \le 1$. Consequently, the transmit beamforming optimization problem with the specified  reflecting beamforming is non-convex. To handle this issue, some prior works\cite{cheng2024networked,10086626,10153696,9916163,hua2024near,9771801,10279464,10440056,10497119,10254508,9124713,10050406,9979782,10496515} proposed the use of the SDR technique, by first relaxing the rank one constraint, i.e., $\mathrm{rank}(\mathbf w\mathbf w^H) \le 1$, to construct an SDP problem. After obtaining the optimal transmit beamforming solution that may be of high-rank, an equivalently optimal rank-one solution is reconstructed for the original transmit beamforming optimization problem\cite{cheng2024networked,10086626,10153696,9916163,hua2024near,9771801,10279464,10440056,10497119,10254508,9124713,10050406,9979782,10496515}. Next, the reflective beamforming design problem in ISAC is also more complex than that in sensing system, due to the non-convex property of the communication SNR. By applying the SCA method to optimize the reflective beamforming design and adopting the AO method to optimize the transmit and reflective beamforming iteratively, a converged beamforming solution is obtained \cite{10440056}.

\subsection{Performance Evaluation}
\begin{figure}[ht]
        \centering
        \includegraphics[width=0.5\textwidth]{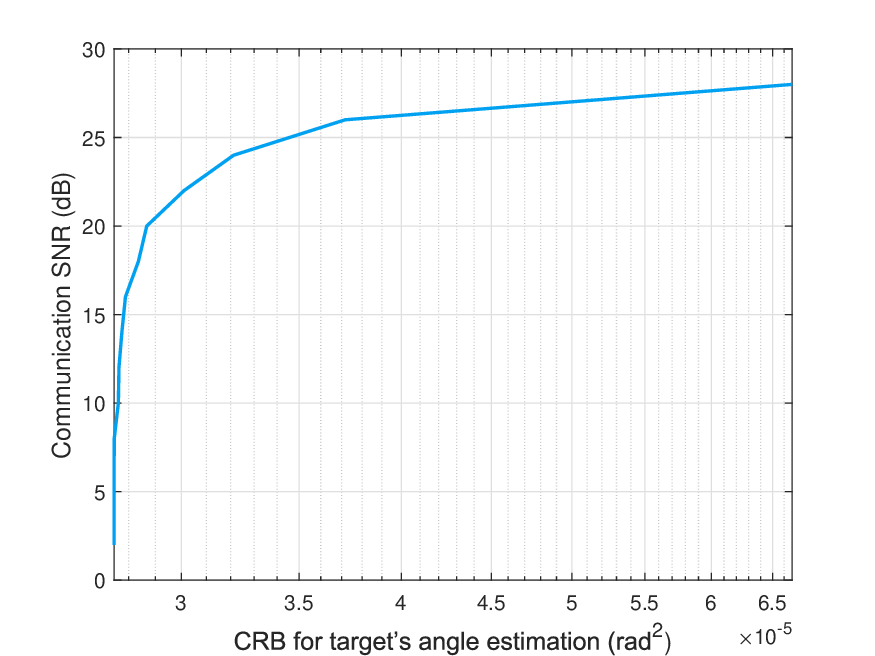}
        \caption{The SNR-CRB region of fully-passive IRS-enabled ISAC.}
        \label{CRB_proposed_point}
\end{figure}

Next, we present the performance tradeoff between CRB for target's angle estimation and communication SNR. We consider a fully-passive IRS-enabled ISAC with one multi-antenna BS, one fully-passive IRS, one CU, and one point target located at the NLoS area of the BS, where the CU can cancel the dedicated sensing signal interference \cite{10440056}. 
For the BS-IRS, IRS-target, and IRS-CU links, the path loss exponent $\alpha_0$ is set as $2.2$, $2.2$, and $3.0$, respectively. The coordinates of the BS and the IRS are  $(0~\text{m},0~\text{m})$ and $(4~\text{m},5~\text{m})$, respectively\cite{10440056}. The CU is randomly located at a rectangular grid with corners $(40~\text{m},0~\text{m})$, $(40~\text{m},-10~\text{m})$, $(50~\text{m},-10~\text{m})$, and $(50~\text{m},0~\text{m})$. The channels of the BS-IRS, IRS-CU, and BS-CU links are modeled as Rician fading channel with the Rician factor being $0.5$\cite{10440056}. Meanwhile, an additional shadow fading is considered for the direct link from the BS and the CU, with a standard deviation value of $10~\text{dB}$\cite{9771801,10440056}. We also set $M_t=M_r=8$, $N=8$, $T=256$, $P_\text{BS} =30~\text{dBm}$, $\sigma^2 = -110~\text{dBm}$, and $\sigma_c^2 = -80~\text{dBm}$\cite{10440056}. 
Under this setup, the communication SNR-CRB region is given in Figure~\ref{CRB_proposed_point}. It is observed that the communication SNR increases with an increase in estimation CRB, which shows the performance tradeoff between communication and sensing.  

\section{Multi-IRS Networked Sensing and Communications}
Despite its advancement, deploying a single IRS-enabled ISAC exhibits some limitations. On the one hand, it is unable to provide a satisfactory coverage performance when there are numerous blockages in the environment. On the other hand, a single IRS merely provides a single view of the target, thereby restricting the sensing performance and potential applications. 

\begin{figure}[ht]
        \centering
        \includegraphics[width=0.9\textwidth]{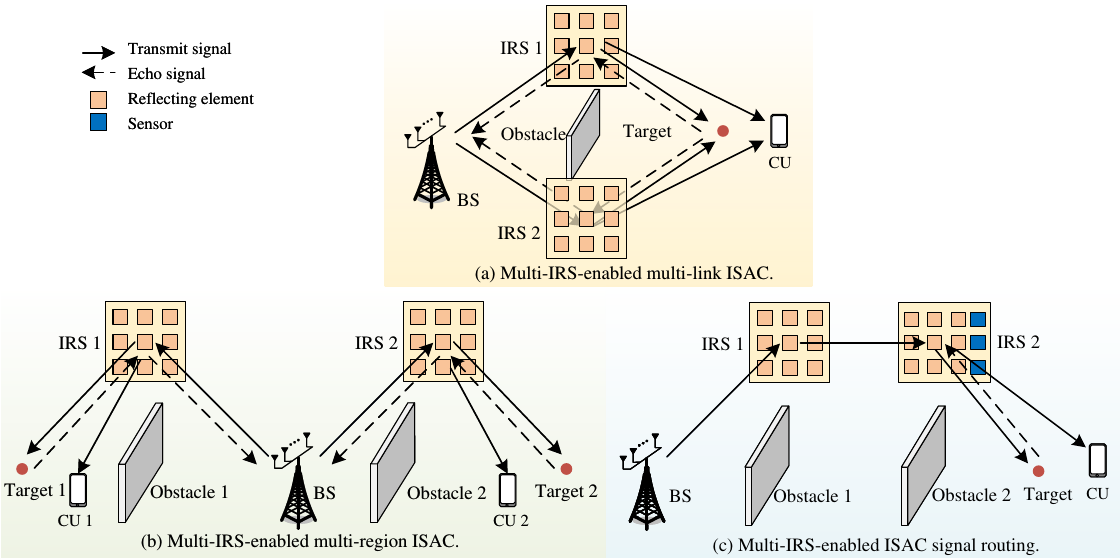}
        \caption{Example scenarios  of multi-IRS-enabled networked ISAC.}
        \label{Networked_ISAC}
\end{figure}
As shown in Figure~\ref{Networked_ISAC}, with the aid of communication ability among different IRSs, multi-IRS-enabled sensing systems are able to form a sensing network to extend the sensing coverage or cooperate with each other to enhance the sensing performance by jointly processing sensing results from multiple angles \cite{10436573,10497119}. As shown in Figure~\ref{Networked_ISAC}(a), multiple IRSs offer a wider range of sensing angles, also known as multi-view sensing. This special feature allows for enhanced perception with, e.g.,  high-accuracy positioning. Furthermore, as shown in Figure~\ref{Networked_ISAC}(b), multiple IRSs can establish multiple virtual LoS links, significantly expanding the coverage areas of both S\&C. Finally, as shown in Figure~\ref{Networked_ISAC}(c), the reflection between multiple IRSs establishes a multi-hop cascaded LoS link between transceivers, which facilitates S\&C in an environment with dense obstacles.

\subsection{Multi-IRS-Enabled Multi-Link ISAC}
As shown in Figure~\ref{Networked_ISAC}(a), cooperative multi-IRS-enabled ISAC is a promising method to enhance sensing accuracy and communication robustness by extending the channel links between the BS and the target/CU. By deploying multiple IRSs at proper locations, the BS transmits signals to the sensing target/CU through the reflection of multiple IRSs. On the one hand, multiple IRSs are able to collaborate with each other to sense a single target from multiple views\cite{fang2024joint}. By jointly exploiting the spatial diversity from multiple views, the sensing performance such as localization accuracy and sensing resolution can be improved. On the other hand, the cooperation between multiple IRSs can also enhance communication performance by improving the transmission link quality between transceivers.

In general, there are two types of collaborative sensing architectures based on how data is aggregated and processed, namely the centralized and distributed cooperative sensing architectures\cite{1458287}. When the multiple IRSs are fully-passive without sensors, only the BS has the ability to receive and process the target’s echo signal. Thus, the sensing data through different links are centralized processed in the BS through the reflection of multiple IRSs. In this case, these multi-view echo signals are superimposed on each other, requiring a complicated and computationally involved sensing algorithm to extract multi-view targets' information. Furthermore, when the distributed IRSs are equipped with sensors, the BS and multiple IRSs all have the ability to receive the target's echo signal, which corresponds the distributed cooperative sensing architecture. In this case, the target’s echo signals are first processed by each IRS. After that, the sensing results after signal processing are further transmitted to the central node (such as the BS) or shared with the neighbor nodes (such as the IRSs). Finally, the central or neighbor nodes jointly process the sensing results collected from various nodes to obtain the global sensing results. Compared with the centralized cooperative sensing architecture, distributed cooperative sensing architecture requires each IRS to have the ability to receive and process the target’s echo signals\cite{10422881}, which inevitably increases the hardware complexity and energy consumption at IRSs.

\subsection{Multi-IRS-Enabled Multi-Region ISAC}
As shown in Figure~\ref{Networked_ISAC}(b), the coverage areas of a single IRS-enabled sensing system are limited by the distance-product path loss and multiple environmental obstacles. To provide a ubiquitous ISAC service, one potential method is to deploy multiple IRSs at proper locations to extend the coverage of the BS\cite{10497119}. In this case, multiple IRSs are deployed at different locations such that each IRS provides S\&C functionalities at one separate region. To balance the performance at multiple regions, the BS transmits multi-beam signals to various IRSs. Meanwhile, as the sensing targets and the CUs in a given region are associated with a single BS, it is necessary but challenging to coordinate the reflective beamforming at all IRSs together with the transmit beamforming at the BS to balance the ISAC performance. 

When the  multiple IRSs are fully-passive, the BS may also receive the echo signals from other areas in addition to the target echo signals from the interested region. This interference between multiple IRSs may decrease the target sensing performance. Thus, advanced sensing algorithms are essential to obtain the target's information in the presence of multi-region interference. Despite mitigating the multi-region interference through beamforming optimization, another possible method is to sense the targets at different regions in a time division manner by controlling the switch of IRSs. Given these above challenges, it is preferable to employ semi-passive or active IRSs by deploying sensors at each IRS to receive the echo signal from each region\cite{10497119}. In this case, due to the obstacles between each region, the interference across different areas can be mitigated. 

\subsection{Multi-IRS-Enabled ISAC Signal Routing}
As shown in Figure~\ref{Networked_ISAC}(c), another application of multi-IRS collaboration is to leverage the multi-reflection of multiple IRSs to establish virtual multi-hop end-to-end links between the BS and the sensing target/CU in an environment with dense obstacles. In this case, deploying multiple IRSs can facilitate the multi-hop signal transmission to strategically bypass potential obstacles. This multi-hop signal transmission link does experience increased path loss, but also leads to more reflective beamforming gains that can be exploited. For instance, prior works \cite{9241752,10159017,10643789} have shown that the multi-IRS-enabled beam routing can significantly enhance the wireless communication performance. Thus, the signal routing between multiple IRSs is promising to enhance the sensing performance. However, unlike wireless communication systems focusing on unidirectional signal transmission between transceivers, the wireless sensing system must collect the target's echo signals. To decrease the huge transmission path loss from the target to the receiver, it is better to deploy sensors at the IRS to collect the echo signals. In multi-IRS-enabled ISAC signal routing systems, one crucial problem is to design the optimal beam routing path from the BS to the sensing target/CU to enhance the channel gain and mitigate the inter-path interference. Usually, the path optimization problem is NP-hard \cite{9241752,10159017,10643789}. Thus, obtaining the optimal solution is generally not possible. In this case, graph theory methods \cite{9241752,10159017,10643789} can be used to find efficient solutions.

\section{Future Directions}
To fully unleash the potential of IRS-enabled sensing and ISAC, several open research problems are worthy of further investigation in the future, which are stated as follows.
\begin{itemize}
	\item Sensing-Assisted IRS-Enabled Communication: For IRS-enabled communication systems, normally, the joint beamforming design highly relies on the availability of CSI between transceivers. However, the CSI acquisition problem can be  challenging due to the involved high-dimensional channel matrices/vectors and the passive property of reflecting elements. In this case, sensing can provide angle/location information of scatterers in the environment, which can be exploited to facilitate the CSI construction based on the coupled relation between S\&C\cite{10584287}. Meanwhile, by using the sensing results based on the reflected echo signals from users, the potential signal propagation links between transceivers can be described in advance to realize timely beam tracking\cite{9791349,10226306,10304580}.

    \item	IRS Deployment: While qualitative guidelines exist for IRS deployment\cite{9586067,9963672,9427474}, the optimal deployment strategy remains unclear. Most existing optimization methodologies rely on instantaneous CSI, which, however, is far from practical given that IRSs can hardly relocate once deployed. This necessitates a sophisticated mathematical formulation for IRS deployment problems based on statistical CSI, along with the corresponding optimization algorithms. In addition, taking into account hardware impairments in IRSs and their impacts on ISAC systems presents a compelling and crucial research topic for the practical implementation of IRSs.

    \item	Sensing with IRS-Mounted Targets: Drawing inspiration from the powerful capability of IRSs to manipulate sensing environments, mounting IRSs on targets emerges as a promising strategy to enable efficient and secure sensing\cite{10274514,10443321}. Specifically, IRSs mounted on targets have the potential to enhance and suppress echo signals directed toward legitimate and eavesdropping radars, respectively. However, the IRS's reflection design requires channel information between the target and transceivers to point the echo signal to the receiver. This underscores the need to explore effective methodologies for estimating critical parameters of channels between the target/IRS and radar by deploying dedicated sensors at the IRS.

\item	Near-Field Sensing and ISAC: Utilizing extremely large multiple-input and multiple-output (XL-MIMO) at mm-Wave/THz bands is an irresistible trend in future 6G systems\cite{hua2024near,10388218,10287779}. This not only leads to an increase in both the antenna-array size and carrier frequency but also signifies a fundamental paradigm shift from conventional far-field communications to its near-field counterparts. However, the coupling of distance and angle in the spherical wave model for near-field wireless channels introduces unique challenges to system design for sensing. There is a need for low-complexity sensing algorithms that can effectively decouple distance and angle while exploiting the characteristics of near-field propagation. Furthermore, hybrid near- and far-field scenarios are worthy of investigation in ISAC, given that both sensing targets and communication users are likely to be in both fields.
\item	IRS-Empowered WideBand ISAC: Wideband ISAC has been increasingly recognized for supporting high data rate transmission and improving sensing resolution  performance. When deploying IRSs to enhance wideband channels for ISAC, the frequency-selective phases at IRSs have to be taken into account. Specifically, the phase response of each IRS element is not flat, but rather, it varies significantly across different frequencies. As a result, the joint design of wideband waveform and frequency-dependent IRS configuration emerges as a crucial aspect in IRS-enabled wideband ISAC systems. Furthermore, the beam squint effect in wideband systems deserves special consideration within the framework of IRS-enabled ISAC.
\item	Machine Learning for IRS-Enabled ISAC Design: Given the complex nature of wireless channels involving IRSs, particularly in multi-IRS-enabled ISAC systems, conventional optimization algorithms either suffer high computational complexity or fail to accurately capture the fundamental properties of such intricate systems. In the context, machine learning-based algorithms stand out as excellent candidates for real-time and lightweight IRS-enabled ISAC design. On the one hand, data-driven deep learning (DL) methods, such as meta-learning and attention mechanisms, can effectively characterize sophisticated system characteristics by leveraging the vast amounts of data collected in wireless networks. On the other hand, model-driven DL, which exploits communication domain knowledge, holds promise for explainable and efficient IRS-enabled ISAC designs.
\end{itemize}
\section{Conclusion}
ISAC has emerged as a pivotal use case for the forthcoming 6G wireless networks. Beyond traditional ISAC work focusing on transceiver design, the deployment of IRS provides new opportunities to control and configure the wireless transmission environment for S\&C performance enhancement. Despite its great potentials, IRS-enabled S\&C faces various unique challenges on architecture design, fundamental sensing and ISAC performance limit analysis, and joint transmit and reflective beamforming design.

To address these issues, this paper presents a timely overview on IRS-enabled S\&C. First, we introduce the basic IRS-enabled sensing architectures with fully-passive, semi-passive, and active IRSs and analyze the fundamental sensing performance limits in terms of detection probability and estimation CRB. Then, we present ISAC signal design and performance boundaries for IRS-enabled ISAC. Furthermore, we introduce the general multi-IRS networked S\&C design framework by sharing and fusing the sensing data from multiple IRSs. Finally, we outline potential research directions in this area for future work.

\bibliographystyle{scm}
\bibliography{SCIS}



\end{document}